\documentclass[amsmath,amssymb,amsbsy,prl,twocolumn,showpacs]{revtex4-1}

\usepackage[pdftex]{graphicx}
\usepackage{epstopdf}
\usepackage{rotating}
\usepackage{dcolumn}
\usepackage{bm}
\usepackage{amsmath}
\usepackage{mathrsfs}
\usepackage{color}
\usepackage{leftidx}
\usepackage{mathtools}

\usepackage[breaklinks,colorlinks=true,linkcolor=blue,urlcolor=blue,citecolor=blue]{hyperref}

\newcommand{\slantedparallel}{\mathbin{\!/\mkern-5mu/\!}}
\begin{document}

\title{Topological magnons with nodal-line and triple-point degeneracies: \\Implications for thermal Hall effect in pyrochlore iridates}

\author{Kyusung Hwang}
\author{Nandini Trivedi}
\author{Mohit Randeria}
\affiliation{Department of Physics, The Ohio State University, Columbus, OH 43210, USA}

\date{\today}

\begin{abstract}
We analyze the magnon excitations in pyrochlore iridates with all-in-all-out (AIAO) antiferromagnetic order, focusing on
their topological features. We identify the magnetic point group symmetries that protect the nodal-line band crossings and 
triple-point degeneracies that dominate the Berry curvature. We find three distinct regimes of magnon band topology, 
as a function of the ratio of Dzyaloshinskii-Moriya (DM) interaction to the antiferromagnetic exchange.
We show how the thermal Hall response provides a unique probe of the topological magnon band structure in AIAO systems.
\end{abstract}

\maketitle

\noindent \underline{\it Introduction:} 
Recently there has been an explosion of activity exploring
topological features in the electronic excitations of semi-metallic and conducting solids. 
This includes the study of Weyl fermions in systems that break either time reversal or inversion symmetry~\cite{Wan2011, Burkov2011, Yang2011,  Xu2015Weyl, Huang2015, Weng2015, Lv2015, Vafek2014,Armitage2018}.
Weyl points act as a source/sink for the Berry curvature in the bulk band structure, and lead to
striking predictions of Fermi arc surface states and the chiral anomaly.
Dirac fermions are realized by fourfold degenerate band crossings protected by time reversal and 
crystal symmetries \cite{Wang2012, Wang2013, Yang2014_DSM, Kargarian2016, Liu2014, Xu2015Dirac, Liu2016}.
Recently discovered triple-point semimetals~\cite{Zhu2016,Chang2017,Weng2016,Weng2016_2,Bradlyn2016,Lv2017}, with triply degenerate band crossings, 
are condensed matter examples of new types of fermions, beyond Weyl and Dirac, with no counterpart in high energy physics.
Nodal-line semimetals are another class of topological systems in which band crossings occur along closed lines in momentum space \cite{Burkov2011NLSM, Fang2015NLSM, Chen2015NLSM, Chen2016NLSM, Bian2016NLSM}.

Topological semimetal band structures are not restricted to fermionic systems and can also arise in bosonic systems.
Spin-orbit coupled magnetic insulators are good candidates to look for ``bosonic" topological semimetals.
In recent studies, a variety of topological features have been predicted in the magnon bands of Cr-based breathing pyrochlore antiferromagnets \cite{Li2016},
pyrochlore ferromagnet Lu$_2$V$_2$O$_7$ \cite{Mook2016}, as well as other magnetic insulators \cite{Fransson2016, SeKwon2016, Owerre2016_2017_honeycomb, Okuma2017, Su2017pyrochlore, Su2017honeycomb, Li2017_3Dhoneycomb, Li2017_Cu3TeO6, Owerre2018_3Dkagome, Zyuzin2018, Jian2018, Owerre2018}.

\begin{figure}[b!]
\centering
\includegraphics[width=\linewidth]{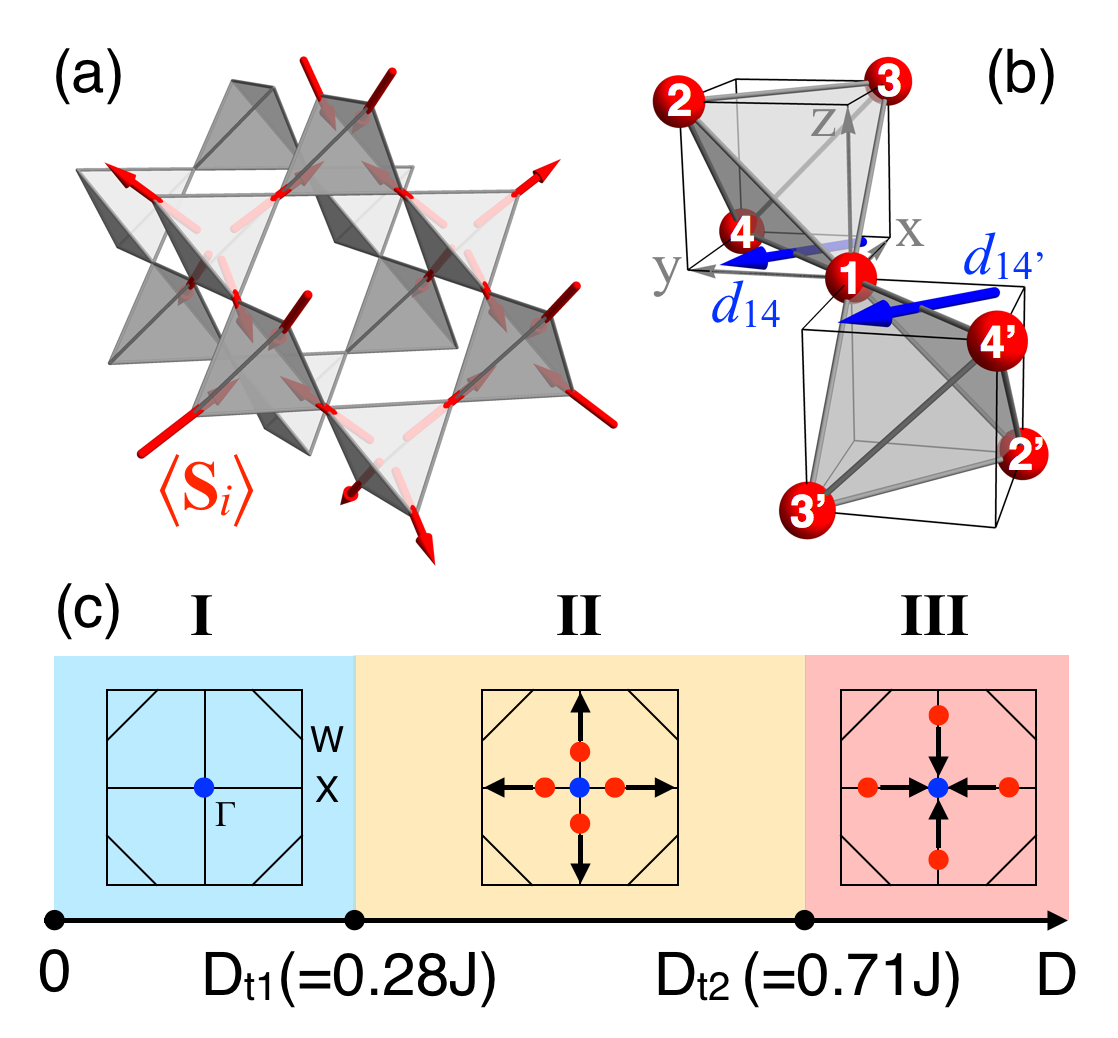}
\caption{
(a) Spin configuration of the AIAO state on the pyrochlore lattice.
(b) Nearest-neighbor DM vectors (blue), $\hat{d}_{14}=\hat{d}_{14'}=(-1,1,0)/\sqrt{2}$.
The unit-length DM vectors $\{\hat{d}_{ij}\}$ are determined by lattice symmetry \cite{Elhajal2005}.
The numbers (1-4) denote the four sublattices of the pyrochlore lattice.
(c) Three regimes (I,II,III) of magnon band topology separated by transition points
$D_{t1}$ and $D_{t2}$, showing magnon triple-points (blue: A-type, red: B-type) in the $k_z=0$ plane of the Brillouin zone [Fig. \ref{fig:magnon_bands} (d)].
Black arrows indicate the directions in which the triple-points move with increasing $D$.
}
\label{fig:pyrochloreAIAO}
\end{figure}

\begin{figure}
\centering
\includegraphics[width=\linewidth]{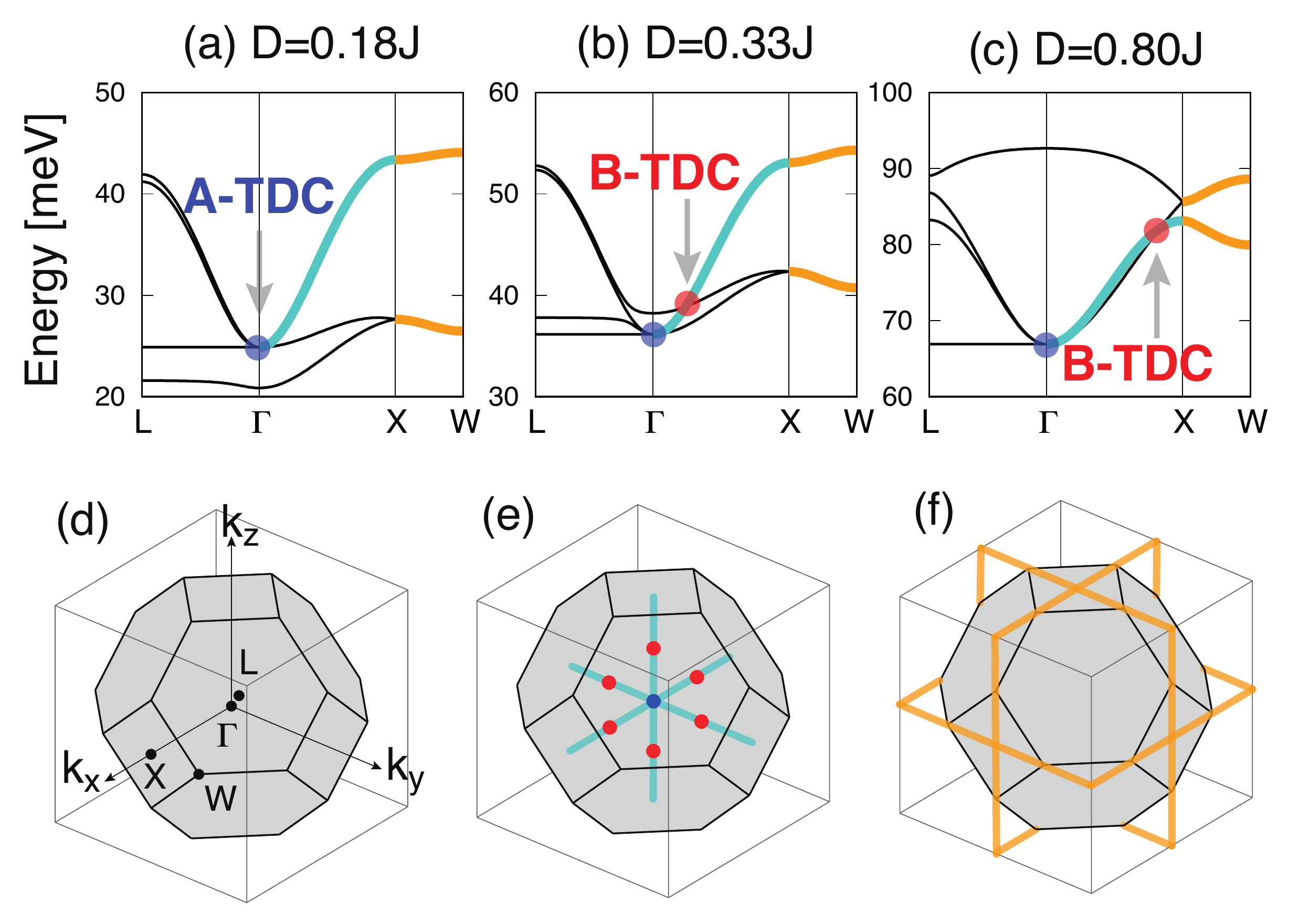}
\begin{ruledtabular}
    \centering
    \begin{tabular}{lcc}
    Band crossing & Location & Symmetry protection \\
    \hline
    A-TP (blue) & $\Gamma$ & $\{C_3,C_2\}$ \\
    B-TP (red) & $\Gamma$X & Inverted bands at $\Gamma$ $\&$ $\Theta*S_4$ \\
    Nodal-line (cyan) & $\Gamma$X & $\Theta*S_4$ \\
    Nodal-line (orange) & XW & $\{G_d,C_2\}$
    \end{tabular}
\end{ruledtabular}
\caption{
Magnon band structures
for (a) $D=0.18J$, (b) $0.33J$, and (c) $0.8J$.
Two types of the triply degenerate band crossings are marked by blue (A-type TDC) and red (B-type TDC) dots.
The cyan line denotes the doubly degenerate band protected by the $\Theta *S_4$ anti-unitary symmetry.
The other doubly degenerate bands along XW (orange) correspond to the nodal-line band crossings protected by the nonsymmorphic $G_d$ glide and $C_2$ rotation.
(d) Brillouin zone of the pyrochlore lattice with high symmetry points [$\Gamma:(0,0,0),~\textup{X}:(\pi/2,0,0),~\textup{W}:(\pi/2,\pi/4,0),~\textup{L}:(\pi/4,\pi/4,\pi/4)$].
(e-f) Locations of the triple-points (e) and nodal-lines (f) in the Brillouin zone, with the same color code as in (a-c).
The table summarizes the associated symmetries protecting the band crossings.
See Supplemental Material \cite{suppl} for the definitions of the symmetry operations.
}
\label{fig:magnon_bands}
\end{figure}

In this work, we propose that the magnon excitations of pyrochlore iridates R$_2$Ir$_2$O$_7$ (R: rare earth or yttrium) \cite{Witczak-Krempa2014,Schaffer2016} 
exhibit triple-point and nodal-line band crossings with unique signatures in thermal Hall effect.
Many pyrochlore iridates are insulators with the all-in-all-out (AIAO) antiferromagnetic order [Fig.~\ref{fig:pyrochloreAIAO}(a)] below $T_c\sim120$~K \cite{Yanagishima2001, Taira2001, Fukazawa2002, Matsuhira2007, Hasegawa2010, Machida2010, Sakata2011, Zhao2011, Matsuhira2011, Tomiyasu2012, Disseler2012, Disseler2012_II, Tafti2012, Shapiro2012, Ishikawa2012, Sagayama2013, Guo2013, Kondo2015, Ueda2015, Tian2016, Clancy2016, Donnerer2016, Wan2011, Witczak-Krempa2012, Go2012, Moon2013, Chen2012, fujita2015, fujita2016, fujita2016_2nd, Lee2013, yamaji2014, hu2012, Chen2015, hu2015, Laurell2017, Yang2014_WSM, Hwang2016, Zhang2017, Wang2017}.
Their spin excitations are relatively less explored experimentally due to neutron absorption by Ir.
Focusing on compounds with nonmagnetic rare earth ion on the R-site, such as Eu$_2$Ir$_2$O$_7$ and Y$_2$Ir$_2$O$_7$, we 
investigate the magnon excitations in the AIAO state 
described by the magnetic interactions between the $S=1/2$  Ir moments on the pyrochlore lattice: 
\begin{equation}
H=\sum_{\langle ij \rangle} \left[ J {\bf S}_i \cdot {\bf S}_j + D\hat{d}_{ij} \cdot {\bf S}_i \times {\bf S}_j\right ] .
\label{eq:model}
\end{equation}
We consider antiferromagnetic (AFM) Heisenberg  exchange $J>0$ and Dzyaloshinskii-Moriya (DM) interactions $D$ between nearest-neighbor moments that
are relevant for AIAO ordering
(ferromagnetic exchange is relevant for Lu$_2$V$_2$O$_7$ \cite{Onose2010,Ideue2012,Mook2016}).
We find two topological transitions in the magnon spectrum with increasing the $D/J$ as shown in Fig.~\ref{fig:pyrochloreAIAO}~(c). 
Interestingly, the three regimes (I,II,III) can be distinguished by their distinct magnon band topology: the triply degenerate crossings of magnon bands, protected by the magnetic point group symmetry of the AIAO state, and
nodal-lines of doubly degenerate band crossings protected by either nonsymmorphic or anti-unitary symmetries.
The degeneracies at the triple-points and nodal-lines make strong contributions to the Berry curvature, which in turn
impact the thermal Hall effect (THE) \cite{Katsura2010,Onose2010,Ideue2012,Matsumoto2011,Matsumoto2014,Hirschberger2015,Lee2015},
an important experimental probe of magnon band topology.

\medspace 

\noindent \underline{\it Model and spin wave theory:}
For the AFM pyrochlore described by Eq.~(\ref{eq:model}), 
the DM interaction plays an important role in selecting the ground state from the highly degenerate ground state manifold in the Heisenberg limit.
We focus on $D>0$ (direct DM), where the ground state is AIAO, whereas for $D<0$ (indirect DM) the ground state has XY 
order ~\cite{Elhajal2005,Witczak-Krempa2012}.
In the AIAO state relevant to the pyrohclore iridates, the spin moments point inward at one type of tetrahedra and outward at the other type 
[Fig.~\ref{fig:pyrochloreAIAO}(a)] with ordering wave vector ${\bf q}={\bf 0}$.

We investigate magnon excitations about the AIAO state using linear spin wave theory.
First, we make a local coordinate transformation for each spin operator that aligns the quantization axis along the moment direction at each site.
Then, we use a linearized Holstein-Primakov transformation to obtain the quadratic Hamiltonian
\begin{eqnarray}
H_{\textup{SW}}=E_{\textup{cl}}
&+& \sum_{\bf k}\sum_{l,m=1}^4 A_{lm}({\bf k})~a_{l{\bf k}}^{\dagger}a_{m{\bf k}}
\nonumber\\
&+& \sum_{\bf k}\sum_{l,m=1}^4 B_{lm}({\bf k})~a_{l-{\bf k}}a_{m{\bf k}} + \textup{H.c.}
\label{eq:H_SW}
\end{eqnarray}
Here $E_{\textup{cl}}$ is the classical ground state energy, and $a^{\dagger}$ ($a$) are the magnon creation (annihilation) operators for
the four magnetic sublattices ($l,m=1,\ldots,4$) of the AIAO state with crystal momentum (${\bf k}$).
The explicit forms of the hopping and pairing amplitudes, $A_{lm}({\bf k})$ and $B_{lm}({\bf k})$, are provided in Supplemental Material~\cite{suppl}.
The corresponding four magnon bands are obtained by diagonalizing $H_{\textup{SW}}$ via the Bogoliubov transformation \cite{suppl}.
We next discuss in turn the two types of topological features in the magnon bands: triple points and nodal lines.

\medspace 

\noindent \underline{\it Triple-points:}
We find that the magnon bands exhibit two types of triply degenerate band crossings (TDC); see Fig.~\ref{fig:magnon_bands}(a).
The first type is a triple degeneracy at the $\Gamma$ point (blue dot) that we denote as the A-type TDC.
It is protected by cubic symmetry ($C_3$ and $C_2$ rotations)~\cite{suppl} and exists irrespective of the size of $D$.

At $D=D_{t1}~(\equiv J\sqrt{2}/5=0.28J)$ there is a band inversion at the $\Gamma$ point between the triply degenerate level and nondegenerate level, resulting in the creation of a second B-type TDC [red dot in Fig.~\ref{fig:magnon_bands}(b)]. 
The B-type TDC arises from the crossing between a nondegenerate and doubly degenerate (cyan) bands along the 6 cubic directions, e.g., $\Gamma$X line.
The double degeneracy of the latter is guaranteed by the anti-unitary symmetry $\Theta *S_4$ of the magnetic point group of the AIAO state~\cite{suppl}.
The B-type  triple-points move toward X and other symmetry related points as $D$ increases.
At $D=D_{t2}~(\equiv J/\sqrt{2}=0.71J)$ another band inversion arises at the X point.
In this band inversion the TDC migrate to the bottom three bands from the top three with the inverted movement direction toward the $\Gamma$ point 
[compare Fig.~\ref{fig:magnon_bands} (c) with (b)].
During this process, a pair of triple-points meet at the X point and then they pass through each other without being annihilated,
due to the different quantum numbers of $(\Theta *S_4)^2$ in the degenerate band ($-1$) and in the other two nondegenerate bands ($+1$) along the $\bf k$ line.

To summarize, the AIAO antiferromagnetic pyrochlore has three regimes (I,II,III) of magnon band topology,
separated by the topological transitions at $D_{t1}$ and $D_{t2}$; see Fig.~\ref{fig:pyrochloreAIAO} (c). 
The magnon band structure in each region is characterized by the pattern of triple-points and their movement in the Brillouin zone.
Note that the AIAO ground state remains stable while the magnon band structure undergoes these topological changes driven by the DM interaction~\cite{suppl}.

\medspace 

\noindent \underline{\it {Nodal-lines:}}
Another characteristic feature of the magnon bands is the existence of nodal-line band crossings. 
Along $\Gamma$X (cyan in Fig. \ref{fig:magnon_bands}) there is a doubly degenerate nodal-line band crossing.  A more interesting nodal-line crossing occurs along the XW and other symmetry-related lines, where four magnon bands are paired up into two doubly degenerate bands [orange in Figs. \ref{fig:magnon_bands} (a-c)] by the symmetry protection of $C_2$ rotation and ``nonsymmorphic" $G_d$ glide~\cite{suppl}. Acting on the magnon operators, these symmetry operations anticommute with each other, resulting in the double degeneracy. Figs. \ref{fig:magnon_bands} (e,f) illustrate both kinds of nodal-lines in the entire Brillouin zone.

We have examined the influence of symmetry-allowed further-neighbor interactions, beyond those included in
Eq.~\eqref{eq:model}, that may exist in real materials. The essential features like the triple-points and nodal-lines are all preserved by symmetry.
Thus their effects on the Berry curvature and the important qualitative features of THE described below all persist,
even in the presence of these further-neighbor interactions~\cite{suppl}.

\begin{figure*}
\centering
\includegraphics[width=0.9\linewidth]{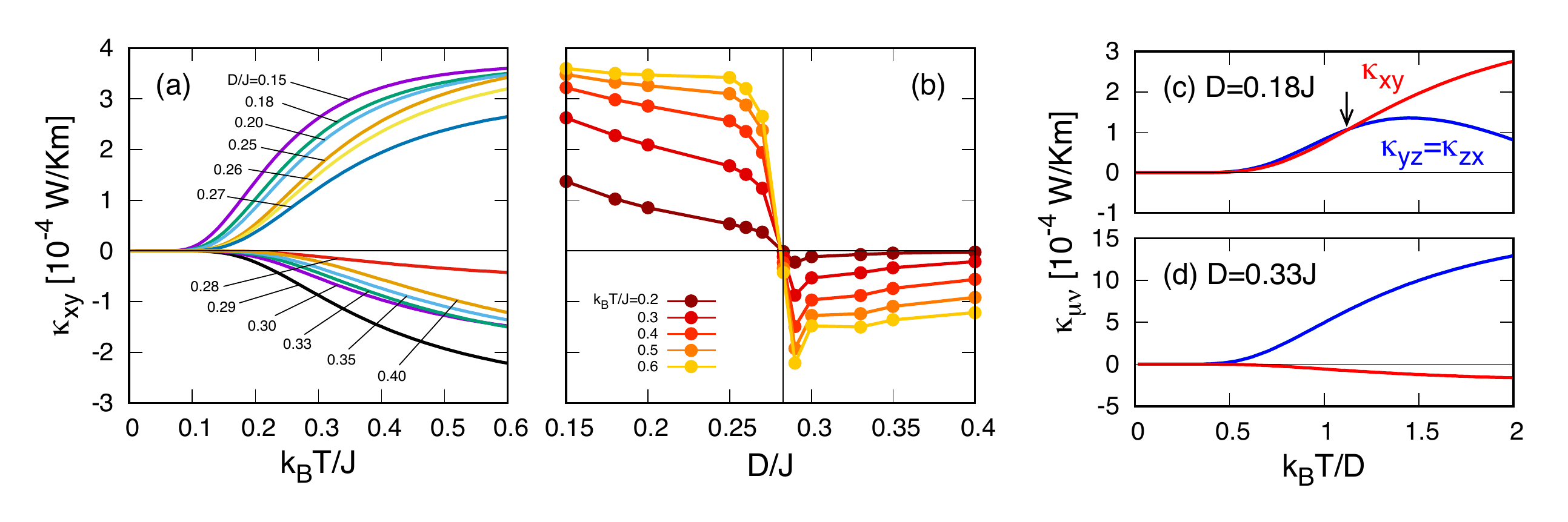}
\caption{
Thermal Hall conductivity $\kappa_{\mu\nu}$ under a small magnetic field $h=0.02J$ along the [110] direction.
(a,b) $\kappa_{xy}$ as a function of $T$ and $D$. The vertical line in (b) corresponds to $D_{t1}.$
(c,d) $\kappa_{xy}$ (red) and $\kappa_{yz}=\kappa_{zx}$ (blue) for $D=0.18J$ and $D=0.33J$.
The arrow in (c) indicates the crossing between $\kappa_{xy}$ and $\kappa_{yz}$ around the temperature scale corresponding to $D$.
}
\label{fig:THC-110field}
\end{figure*}

\medspace

\noindent \underline{\it Berry curvature and thermal Hall effect:}
The distinct magnon band topology exhibited in the three regimes leads to qualitatively different patterns of Berry curvature in the band structure of each regime.
A direct experimental signature of the magnon Berry curvature is the magnon thermal Hall effect \cite{Katsura2010,Onose2010,Ideue2012,Matsumoto2011,Matsumoto2014}.
A temperature gradient $\nabla_{\nu} T$ induces transverse heat current $J_{\mu}^Q =-\sum_{\nu}\kappa_{\mu\nu}\nabla_{\nu} T$ 
carried by magnon excitations as a result of their Berry curvature 
$\boldsymbol{\Omega}_{n{\bf k}}=({\Omega}_{n{\bf k}}^x,{\Omega}_{n{\bf k}}^y,{\Omega}_{n{\bf k}}^z)$.
The antisymmetric thermal Hall conductivity tensor $\kappa_{\mu\nu}$ obtained from linear response theory is given by \cite{Matsumoto2014}
\begin{equation}
\kappa_{xy}
=\frac{k_B^2 T}{\hbar V}
\sum_{n=1}^4 \sum_{\bf k} 
\left\{
\frac{\pi^2}{3} - c_2[g(E_{n{\bf k}}/{k_BT})] 
\right\}
\Omega_{n{\bf k}}^z.
\label{eq:THC}
\end{equation}
$\kappa_{yz}$ and $\kappa_{zx}$ are obtained by cyclic permutations of indices.
Here $c_2(u)=(1+u)\left( \textup{ln} \frac{1+u}{u} \right)^2-(\textup{ln}u)^2-2\textup{Li}_2(-u)$, with $\textup{Li}_2(x)$ the dilogarithm function, 
$g(x)=(e^{x}-1)^{-1}$ is the Bose distribution, $E_{n{\bf k}}$ the magnon dispersion, and $V$ the volume of the system.
For each magnon band, the Berry curvature is given by~\cite{suppl,Matsumoto2014}
\begin{equation}
\boldsymbol{\Omega}_{n{\bf k}} = i \sum_{m=1}^8 \frac{\partial [T_{{\bf k}}^{-1}]_{nm}}{\partial {\bf k}} \boldsymbol{\times}  \frac{\partial [T_{{\bf k}}]_{mn}}{\partial {\bf k}},
\label{eq:Berry}
\end{equation}
where $T_{\bf k}$ is the $8\times 8$ Bogoliubov transformation matrix corresponding to
the four bands and two ``particle-hole" degrees of freedom.

\begin{table}[b]
\begin{ruledtabular}
\begin{tabular}{ccc}
${\bf h}$ & ${\bf M}=(M_x,M_y,M_z)$ & $\boldsymbol{\kappa}=(\kappa_{yz},\kappa_{zx},\kappa_{xy})$
\\
\hline
$[100]$ & $M_y=M_z=0$ & $\kappa_{zx}=\kappa_{xy}=0$
\\
$[110]$ & $M_x=M_y$ & $\kappa_{yz}=\kappa_{zx}$
\\
$[111]$ & $M_x=M_y=M_z$ & $\kappa_{yz}=\kappa_{zx}=\kappa_{xy}$
\end{tabular}
\end{ruledtabular}
\vspace{-0.2cm}
\caption{Induced magnetization for several field directions, and resulting symmetry constraints on the thermal Hall conductivity tensor.
The constraint for the [110] field direction holds for intermediate field directions between [110] and [111], \it{i.e.,} $\hat{h}= \frac{1}{\sqrt{2}}(\hat{x}+\hat{y}) \textup{cos}\theta+\hat{z}\textup{sin}\theta$.
}
\label{tab:symm-constraints}
\end{table}

\begin{figure*}
\centering
\includegraphics[width=\linewidth]{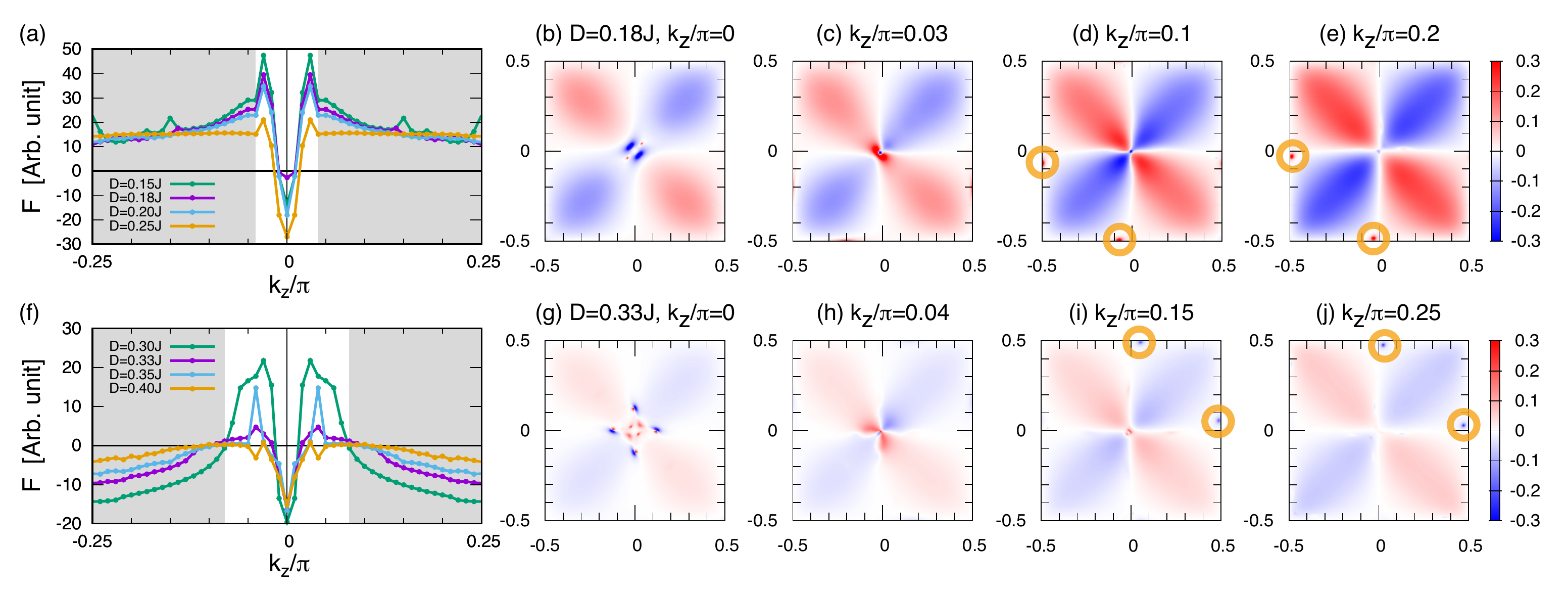}
\vspace{-0.5cm}
\caption{
Momentum-resolved thermal Hall conductivity $K_{xy}({\bf k})$ at $k_B T=0.3J$ under an applied field $h=0.02J\slantedparallel[110]$. Left: $F=\int_{-\pi/2}^{\pi/2} dk_x \int_{-\pi/2}^{\pi/2} dk_y K_{xy} ({\bf k})$ as a function of $k_z$ for (a) regime I and (f) regime II.
Right: Color map of $K_{xy}$ for different $k_z$ slices for (b-e) $D=0.18J$ and (g-j) $D=0.33J$.
In each map, the horizontal and vertical axes represent $k_x/\pi$ and $k_y/\pi$, respectively.
See the text for an explanation of the shaded gray regions and orange circles.
}
\vspace{-0.5cm}
\label{fig:mom-res-THC}
\end{figure*}
\begin{figure}[b]
\centering
\vspace{-0.65cm}
\includegraphics[width=\linewidth]{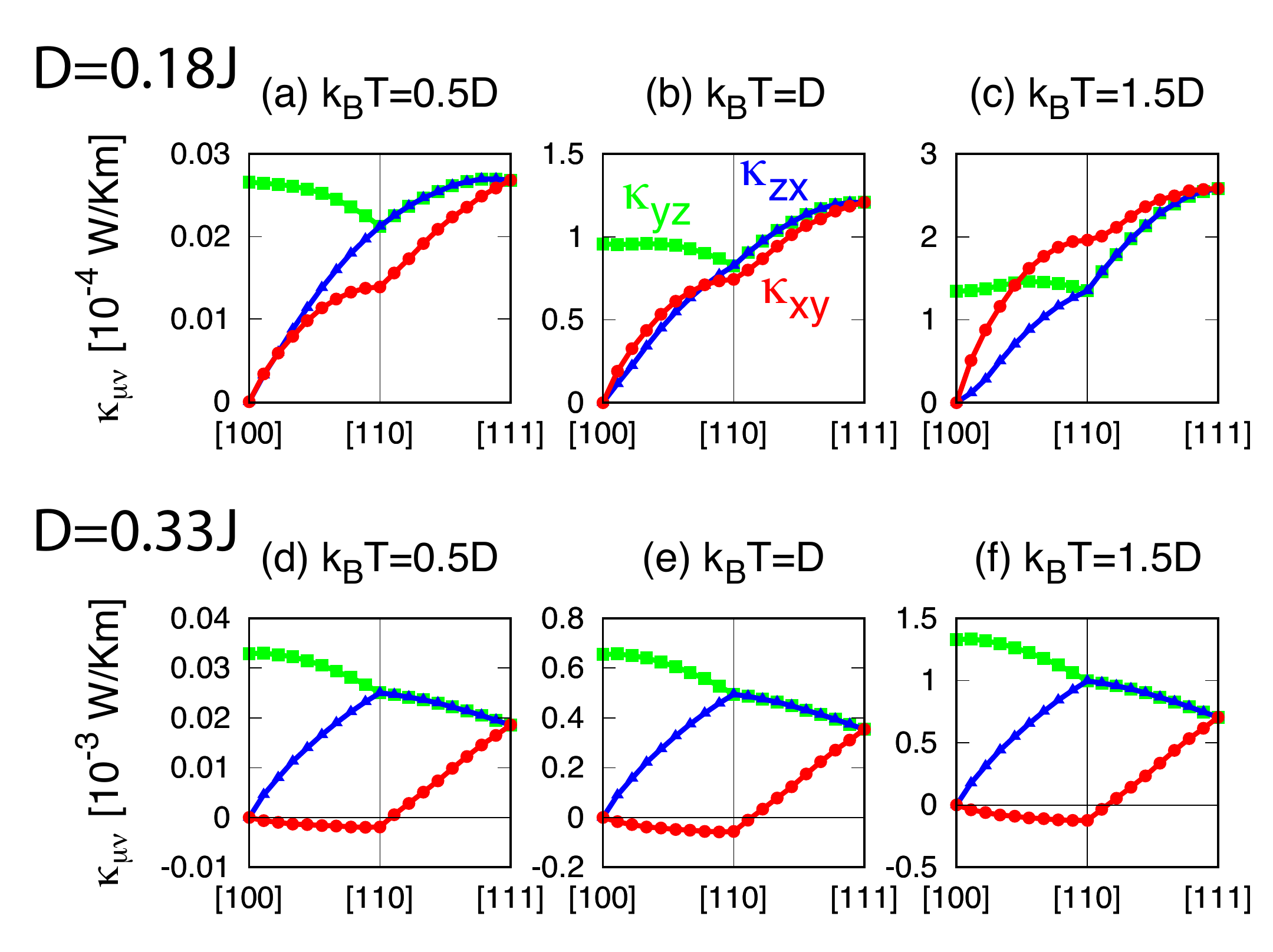}
\caption{
Field-direction dependence of $\kappa_{\mu\nu}$ for $D=0.18J$ (a-c) and $D=0.33J$ (d-f).
The horizontal axis represents the field direction $\hat{h}$ which changes from the [100] to [110] and then [111] direction, with the fixed magnitude $h=0.02J$.
}
\label{fig:THC-field-direction}
\end{figure}

Before discussing our results, we note that the cubic symmetry prohibits finite thermal Hall effect, even though the Berry curvature is 
locally nonzero in the Brillouin zone \cite{Yang2011,Yang2014_WSM,Hwang2016,Laurell2017}.
To probe the magnon Berry curvature via thermal Hall effect, we break the cubic symmetry by applying a small magnetic field to the system.
The Zeeman coupling $H_{\textup{Z}}=-{\bf h}\cdot\sum_i{\bf S}_i$ generates a sublattice-dependent potential in the spin wave Hamiltonian $H_{\textup{SW}}$ as well as canting of the AIAO spin configuration (thereby a nonzero magnetization)~\cite{suppl}.
In Table~\ref{tab:symm-constraints}, we summarize the direction of induced magnetization for several field directions, and also symmetry constraints on the thermal Hall conductivity tensor.
The constraints are obtained based on:
(i) remaining symmetries in the canted AIAO state under the field, and (ii) the fact that the tensor $\boldsymbol{\kappa}=(\kappa_{yz},\kappa_{zx},\kappa_{xy})$ and magnetization ${\bf M}=(M_x,M_y,M_z)(\equiv\frac{1}{4}\sum_{i=1}^{4} \langle {\bf S}_i \rangle)$ over a unit cell, are both axial vectors that follow the same transformation rules under symmetry operations.

To present numerical results for $\kappa_{\mu\nu}$, and to show that their magnitude
is testable by current experiments, we need to estimate the parameters of our model. 
Recent resonant inelastic X-ray scattering experiments~\cite{Donnerer2016} on Sm$_2$Ir$_2$O$_7$ show magnetic excitations 
well described by Eq.~(\ref{eq:model}) with $J\!=\!27.3$ meV and $D\!=0.18J\!=\!4.9$ meV.
We use $J\!=\!27.3$ meV to compute the thermal Hall effect,
and examine our results as a function of $D/J$.

In Fig. \ref{fig:THC-110field}, we show $\kappa_{\mu\nu}$ with a small field $h=0.02J$ along the [110] direction, 
for which the relationship between magnon band topology and thermal Hall response is most clearly observed.
The presence of a small field breaks all the symmetries listed in the table of Fig. \ref{fig:magnon_bands}. Nonetheless, triple-point and nodal-line band crossings remain nearly degenerate carrying large Berry curvatures.
We find characteristic behaviors in the thermal Hall conductivity that can help distinguish the regime I and II in experiments.
Specifically, $\kappa_{xy}$ has a different sign in the two regimes: positive in the regime I and negative in the regime II as shown in 
Fig.~\ref{fig:THC-110field}(a).
In Fig.~\ref{fig:THC-110field}(b) we show that $\kappa_{xy}$ changes sign across the boundary $D_{t1}=0.28J$ (vertical line).
The other two components ($\kappa_{yz}=\kappa_{zx}$) are positive below $T_c$, regardless of which regime the system lies in [Figs. \ref{fig:THC-110field}(c,d)].
We find the same pattern of $\kappa_{\mu\nu}$ even upon inclusion of further neighbor interactions \cite{suppl}.

To get insight about the qualitatively different behavior of $\kappa_{xy}$ in regimes I and II, we resolve it in 
momentum space: $\kappa_{xy}={1\over V}\sum_{\bf k} K_{xy}({\bf k})=\int_{-\pi/4}^{\pi/4} dk_z F(k_z)$.
The $k_z$-variation of the ``integrated" quantity $F(k_z) \equiv \int_{-\pi/2}^{\pi/2} dk_x \int_{-\pi/2}^{\pi/2} dk_y K_{xy} ({\bf k})$,
plotted in the left panels of Fig.~\ref{fig:mom-res-THC}, reveals important features of $\kappa_{xy}$ in momentum space:
(i) a peak structure around $k_z=0$ that changes sign (white), and, (ii) monotonic behavior with no sign-change away from the center (gray).
These plots show that constructive contributions from the gray region determine the sign of $\kappa_{xy}$ in each of the two regimes.

In the right panels of Fig.~\ref{fig:mom-res-THC}, we plot $K_{xy} ({\bf k})$ as a function of $(k_x,k_y)$ for various $k_z$ slices,
to gain a better understanding of how the degeneracies of the (zero-field) magnon spectrum impact the Berry curvature
and hence the sign of $\kappa_{xy}$. The large and rapidly changing behavior of $F(k_z)$ near $k_z=0$ is seen to arise
from the Berry curvature concentrated around the triple-points (b-c,g-h).

More importantly, the doubly degenerate nodal-lines along XW contribute to the large positive (negative) $F(k_z)$ at large $k_z$ 
in regime I (II). The peaks highlighted by circles in the right most panels correspond to the location of the degenerate energy levels (d-e,i-j). 
In a small field, these levels are shifted from XW lines and their degeneracy is lifted, but only slightly, so they continue to give important
contributions to $F(k_z)$.
Notice that in regime I the nodal-lines are shifted into the third quadrant of each $k_z$ plane with positive $K_{xy}({\bf k})$. By contrast, in regime II they move into the first quadrant with negative $K_{xy}({\bf k})$. It is therefore the distinct field-response of the nodal-line topological magnons that ultimately controls the sign of $\kappa_{xy}$ in Fig. \ref{fig:THC-110field}.
The other non-degenerate bands generate the clover-leaf shaped ``background" contributions in the right panels of Fig.~\ref{fig:mom-res-THC}.

From the above analysis, we see that different band topologies lead to distinct patterns of magnon Berry curvature, which in turn 
lead to different thermal Hall responses (see Fig. \ref{fig:THC-110field}), indicating its usefulness 
as a probe of the overall magnon band topology.

The field-direction dependence of $\kappa_{\mu\nu}$ provides additional information about the two regimes, as depicted in Fig. \ref{fig:THC-field-direction}
for two values of $D=0.18J$ and $D=0.33J$. We find: 
(i) $\kappa_{xy}\ge 0$ in regime I but becomes negative along [100] to [110] in regime II. 
(ii) In regime I, $\kappa_{xy}$ and $\kappa_{zx}$ cross as the temperature drops below $D$ regardless of the field direction [Figs. \ref{fig:THC-field-direction} (a-c)].
This generic crossing behavior can be useful for estimating the size of the DM coupling in thermal Hall experiments.

\noindent \underline{\it Conclusions:}
By combining spin wave theory, Berry curvature analysis for the magnon bands, and linear response theory, we have shown how thermal Hall effect can probe the topology of the magnon bands and provide an estimate of the DM coupling. 
One of the central ideas explored here is the field-response of the topological magnon nodal-lines and triple-points and their manifestation in the thermal Hall transport.
Going forward, our calculations suggest that pump-probe techniques that can excite magnons near the doubly and triply degenerate energy levels
will lead to enhanced thermal Hall effect compared to simply relying on thermally excited magnons. 
Our calculations of the thermal Hall transport can also be extended to other iridium-based magnetic insulators \cite{Rau2016}, for example,
pyrochlore iridates with a magnetic rare earth ion (such as  Nd$_2$Ir$_2$O$_7$) that bring in an additional source of magnetism. 

\noindent \underline{\it Acknowledgements:}
We are grateful to Arun Paramekanti and James Rowland for helpful discussions. 
This work was supported by NSF MRSEC grant DMR-1420451 (K.H. and M.R.) and NSF DMR-1309461 (N.T.).



\pagebreak
\begin{center}
\textbf{\Large Supplemental Material}
\end{center}

\section{Magnon Hamiltonian}
\begin{figure}[b]
\centering
\includegraphics[width=0.75\linewidth]{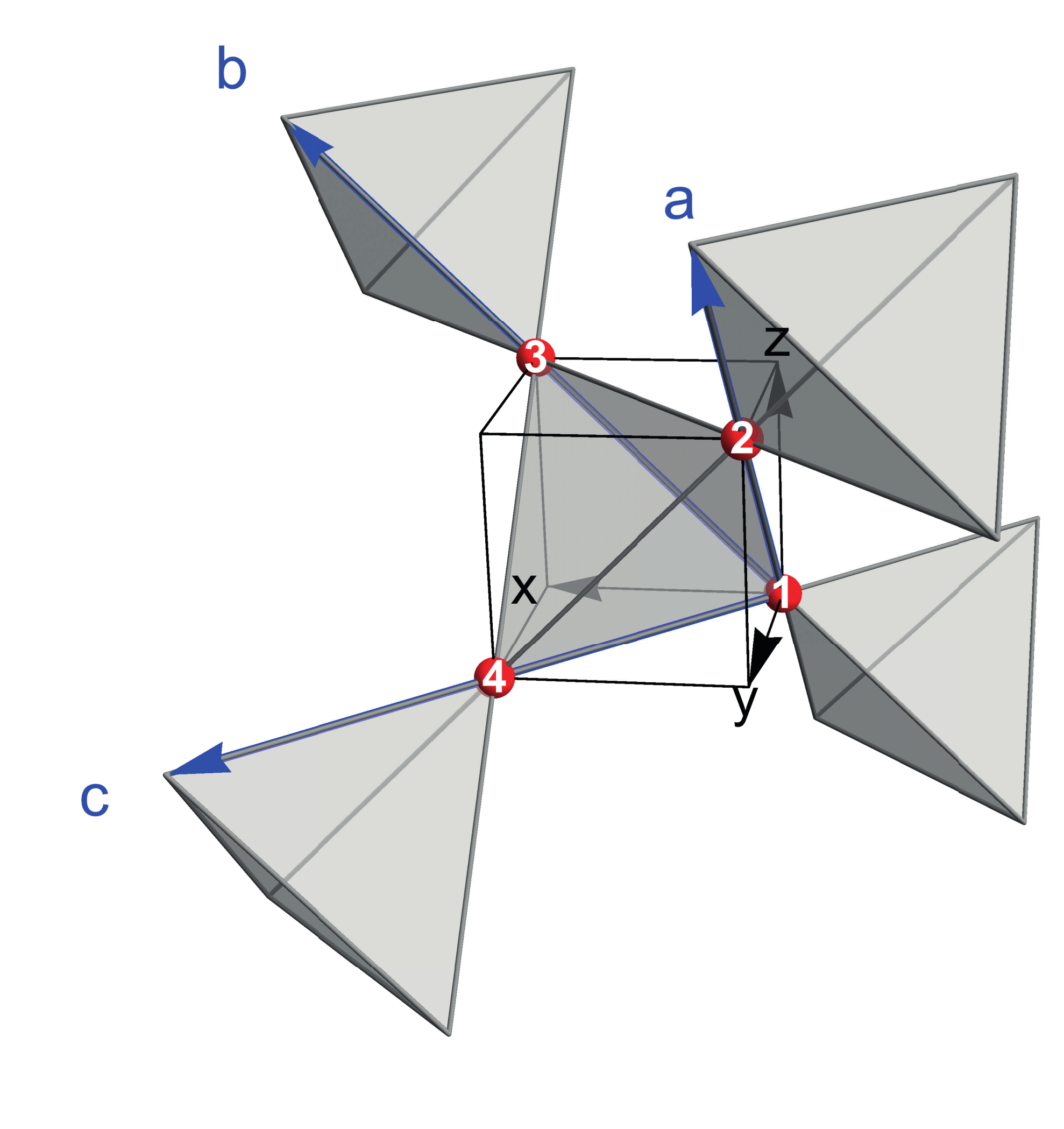}
\caption{Pyrochlore lattice in a cubic frame.
The numbered red balls ($1,\cdots,4$) represent four sublattices in a unit cell, and the blue arrows denote the lattice vectors, ${\bf a}=(0,2,2)$, ${\bf b}=(2,0,2)$, ${\bf c}=(2,2,0)$.
}
\label{fig:structure}
\end{figure}

The all-in-all-out (AIAO) state is a ${\bf q}=0$ magnetic order on the pyrochlore lattice, hence with four magnetic sublattices.
Spin moments on the four sublattices point inward or outward at each local tetrahedron (Fig. 1 of the main text).
In the cubic coordinate system of Fig. \ref{fig:structure} (or Fig. 1), the classical  AIAO spin configuration can be expressed as follows.
\begin{eqnarray}
\textup{AIAO:}~~~~~
&&
{\bf S}_1^\textup{cl} \slantedparallel (\bar{1},\bar{1},\bar{1}),
\nonumber\\
&&
{\bf S}_2^\textup{cl} \slantedparallel (\bar{1},1,1),
\nonumber\\
&&
{\bf S}_3^\textup{cl} \slantedparallel (1,\bar{1},1),
\nonumber\\
&&
{\bf S}_4^\textup{cl} \slantedparallel (1,1,\bar{1}),
\end{eqnarray}
where the subscript means the sublattice.

As a first step to describe magnon excitations in the AIAO state, we define local axes $\{{\bf l}^x,{\bf l}^y,{\bf l}^z\}$ at each sublattice in such a way that the local axis ${\bf l}^z$ is parallel to the moment direction ${\bf S}^\textup{cl}$ (see Table \ref{tab:localaxes_convention} for our local axes convention).
Taking the coordinate transformation ${\bf S}_m \rightarrow \Lambda_m {\bf S}_m$ with the transformation matrix $\Lambda_m\equiv({\bf l}_m^x,{\bf l}_m^y,{\bf l}_m^z)$ for each sublattice $m~(=1,\cdots,4)$,
we rewrite  the spin model in the frame of the local axes:
\begin{equation}
H=\sum_{\langle ij \rangle} S_i^{\alpha} \Lambda_i^{\mu\alpha} \mathcal{J}_{ij}^{\mu\nu} \Lambda_j^{\nu\beta} S_j^{\beta} ,
\end{equation}
where $\mu,\nu,\alpha,\beta\in\{x,y,z\}$, and $\mathcal{J}_{ij}^{\mu\nu} = J \delta_{\mu\nu} + D d_{ij}^{\rho} \epsilon_{\rho\mu\nu}$ with the Kronecker delta $\delta_{\mu\nu}$ and totally antisymmetric tensor $\epsilon_{\rho\mu\nu}$.
We use the Einstein summation convention for repeated Greek indices.

Quadratic magnon Hamiltonian is obtained by applying the linearized Holstein-Primakov representation \cite{Holstein1940} to the above Hamiltonian.
\begin{equation}
\begin{array}{ccl}
{S}^x &=& \sqrt{\frac{S}{2}} (a+a^{\dagger}),
\\
{S}^y &=& -i \sqrt{\frac{S}{2}} (a-a^{\dagger}),
\\
{S}^z &=& S- a^{\dagger} a.
\end{array}
\end{equation}
Spin operators are now written in terms of the boson operators $\{a,a^{\dagger}\}$ and the size of spin $S$.
Large-$S$ expansion of the Hamiltonian followed by Fourier transformation leads to the quadratic magnon Hamiltonian $H_{\textup{SW}}$ in Eq. 2 of the main text, equivalently 
\begin{equation}
H_{\textup{SW}}=E_{\textup{cl}}-\frac{1}{2}  \sum_{m=1}^4 \sum_{{\bf k}} A_{mm}({\bf k})
+ \frac{1}{2} \sum_{\bf k} \Psi^{\dagger}_{\bf k} H_{\bf k} \Psi_{\bf k},
\end{equation}
where $E_{\textup{cl}}=-4(J+2\sqrt{2}D)N_{\textup{uc}}S^2$ ($N_{\textup{uc}}$: the number of unit cells),
$\Psi_{\bf k}=[a_{1{\bf k}} \cdots a_{4{\bf k}}|a_{1-{\bf k}}^{\dagger} \cdots a_{4-{\bf k}}^{\dagger}]^T$, 
and
\begin{equation}
H_{\bf k}
=
\left[
\begin{array}{c|c}
{\bf A}({\bf k}) & 2 {\bf B}^{\dagger}({\bf k})
\\
\hline
2 {\bf B}({\bf k}) & {\bf A}^T({-{\bf k}})
\end{array}
\right].
\end{equation}
The 4$\times$4 matrices ${\bf A}$ and ${\bf B}$ representing the hopping and pairing of the bosons are given by
\begin{widetext}
\begin{eqnarray}
{\bf A}({\bf k})
&=&
2S(J+2\sqrt{2}D)
\left[
\begin{array}{cccc}
1&0&0&0
\\
0&1&0&0
\\
0&0&1&0
\\
0&0&0&1
\end{array}
\right]
\nonumber\\
&-&
\frac{2S}{3}(J-\sqrt{2}D)
\left[
\begin{array}{c|c|c|c}
0 & e^{i\pi/3} p({\bf a}) & p({\bf b}) & e^{i\pi/3} p({\bf c})
\\
\hline
e^{-i\pi/3} p({\bf a}) & 0 & e^{-i\pi/3} p({\bf a}-{\bf b}) & p({\bf a}-{\bf c})
\\
\hline
p({\bf b}) & e^{i\pi/3} p({\bf a}-{\bf b}) & 0 & e^{i\pi/3} p({\bf b}-{\bf c})
\\
\hline
e^{-i\pi/3} p({\bf c}) & p({\bf a}-{\bf c}) & e^{-i\pi/3} p({\bf b}-{\bf c}) & 0
\end{array}
\right],
\\
{\bf B}({\bf k})
&=&
\frac{2S}{6}(2J+\sqrt{2}D)
\left[
\begin{array}{c|c|c|c}
0 & e^{i2\pi/3} p({\bf a}) & e^{-i\pi/3} p({\bf b}) & e^{-i2\pi/3} p({\bf c})
\\
\hline
e^{i2\pi/3} p({\bf a}) & 0 & e^{-i2\pi/3} p({\bf a}-{\bf b}) & e^{i\pi/3} p({\bf a}-{\bf c})
\\
\hline
e^{-i\pi/3} p({\bf b}) & e^{-i2\pi/3} p({\bf a}-{\bf b}) & 0 & e^{i2\pi/3} p({\bf b}-{\bf c})
\\
\hline
e^{-i2\pi/3} p({\bf c}) & e^{i\pi/3} p({\bf a}-{\bf c}) & e^{i2\pi/3} p({\bf b}-{\bf c}) & 0
\end{array}
\right],
\end{eqnarray}
\end{widetext}
where $p({\bf R})=\textup{cos}({\bf k}\cdot{\bf R}/2)$, ${\bf k}=(k_x,k_y,k_z)$, and $\{{\bf a},{\bf b},{\bf c}\}$ are the lattice vectors (Fig. \ref{fig:structure}).
Here we notice that the hopping matrix ${\bf A}$ encodes a fictitious flux pattern that is seen by magnons.
Magnons pick up $\pi$ (0) flux when they hop around a local triangle (hexagon) on the pyrochlore lattice.
The nontrivial flux pattern arises due to the noncoplanar AIAO magnetic structure that is stabilized by the DM interaction.
However, it becomes trivial (zero flux for both triangle and hexagon) when $D>D_{t2}(=J/\sqrt{2})$.
In this AIAO antiferromagnetic pyrochlore, magnons experience quantized fictitious fluxes (0 or $\pi$) at each elementary triangle and hexagon.
This contrasts with the ferromagnetic pyrochlore where fluxes vary continuously as functions of the DM interaction \cite{Onose2010,Ideue2012}.

\begin{table}
\begin{ruledtabular}
\begin{tabular}{ccccc}
Sublattice ($m$) & ${\bf r}_m$ & ${\bf l}_m^z$ & ${\bf l}_m^x$ & ${\bf l}_m^y$
\\
\hline
&&&&
\\
1 & $(0,0,0)$ & $\frac{1}{\sqrt{3}}(\bar{1},\bar{1},\bar{1})$ & $\frac{1}{\sqrt{2}}(1,\bar{1},0)$ & $\frac{1}{\sqrt{6}}(\bar{1},\bar{1},2)$
\\
&&&&
\\
2 & $(0,1,1)$ & $\frac{1}{\sqrt{3}}(\bar{1},1,1)$ & $\frac{1}{\sqrt{2}}(0,1,\bar{1})$ & $\frac{1}{\sqrt{6}}(\bar{2},\bar{1},\bar{1})$
\\
&&&&
\\
3 & $(1,0,1)$ & $\frac{1}{\sqrt{3}}(1,\bar{1},1)$ & $\frac{1}{\sqrt{2}}(\bar{1},\bar{1},0)$ & $\frac{1}{\sqrt{6}}(1,\bar{1},\bar{2})$
\\
&&&&
\\
4 & $(1,1,0)$ & $\frac{1}{\sqrt{3}}(1,1,\bar{1})$ & $\frac{1}{\sqrt{2}}(0,1,1)$ & $\frac{1}{\sqrt{6}}(2,\bar{1},1)$
\\
&&&&
\end{tabular}
\end{ruledtabular}
\caption{Position vectors (${\bf r}$) and local axes (${\bf l}^x,{\bf l}^y,{\bf l}^z$) of the four sublattices shown in Fig. \ref{fig:structure}.}
\label{tab:localaxes_convention}
\end{table}

The magnon Hamiltonian is diagonalized via the Bogoliubov transformation that relates the bare bosons ($a$) with magnon quasiparticle modes ($b$):
\begin{equation}
\Psi_{\bf k}=T_{\bf k} \Gamma_{\bf k}.
\end{equation}
Here $T_{\bf k}$ is the 8$\times$8 Bogoliubov transformation matrix, and $\Gamma_{\bf k}=[b_{1{\bf k}} \cdots b_{4{\bf k}}|b_{1-{\bf k}}^{\dagger} \cdots b_{4-{\bf k}}^{\dagger}]^T$ is a column vector containing magnon quasiparticle modes.
The transformation matrix is obtained by solving the bosonic eigenvalue problem
\begin{equation}
T^{\dagger}_{\bf k} H_{\bf k} T_{\bf k} = E_{\bf k}.
\end{equation}
In this equation, eigenvectors contained in the columns of $T_{\bf k}$ are paired with magnon energy eigenvalues stored in the diagonal matrix $E_{\bf k}=\textup{diag}[E_{1{\bf k}} \cdots E_{4{\bf k}}|E_{1{\bf k}} \cdots E_{4{\bf k}}]$.
It is important to note that due to the boson statistics $T_{\bf k}$ satisfies the para-unitary condition \cite{Matsumoto2014}
\begin{equation}
T^{\dagger}_{\bf k} \sigma_3 T_{\bf k} = \sigma_3 = T_{\bf k} \sigma_3 T^{\dagger}_{\bf k},
\end{equation}
where $\sigma_3=\textup{diag}[1,1,1,1|-1,-1,-1,-1]$ is a 8$\times$8 diagonal matrix that distinguishes the particle and hole sectors of $\Psi$ or $\Gamma$.
The Bogoliubov transformation finally leads to the diagonalized magnon Hamiltonian
\begin{equation}
H_{\textup{SW}}= \textup{constant} + \sum_{n=1}^4 \sum_{{\bf k}} E_{n{\bf k}} b_{n{\bf k}}^{\dagger} b_{n{\bf k}}.
\end{equation}

When a magnetic field is applied to the system ($H_\textup{Z}=-{\bf h}\cdot\sum_i {\bf S}_i$), the above zero-field magnon Hamiltonian $H_{\textup{SW}}$ is modified by two factors: (i) spin moment canting by the magnetic field and (ii) sublattice dependent chemical potential.
The canted spin configuration, which is obtained by solving the spin model $H+H_\textup{Z}$ classically, redefines the local axes $\{{\bf l}^x,{\bf l}^y,{\bf l}^z\}$ at each sublattice, leading to changes in the hopping and pairing amplitudes of $H_{\bf k}$.
The Zeeman coupling $H_{\textup{Z}}$ generates a nonuniform chemical potential term that varies depending on the sublattice and moment direction:
$H_{\textup{Z}} \rightarrow -\sum_{m=1}^4\sum_{\bf k} ({\bf h} \cdot {\bf l}_m^z) a_{m{\bf k}}^{\dagger} a_{m{\bf k}}$.

\section{Symmetry analysis}

\begin{figure}
\centering
\includegraphics[width=\linewidth]{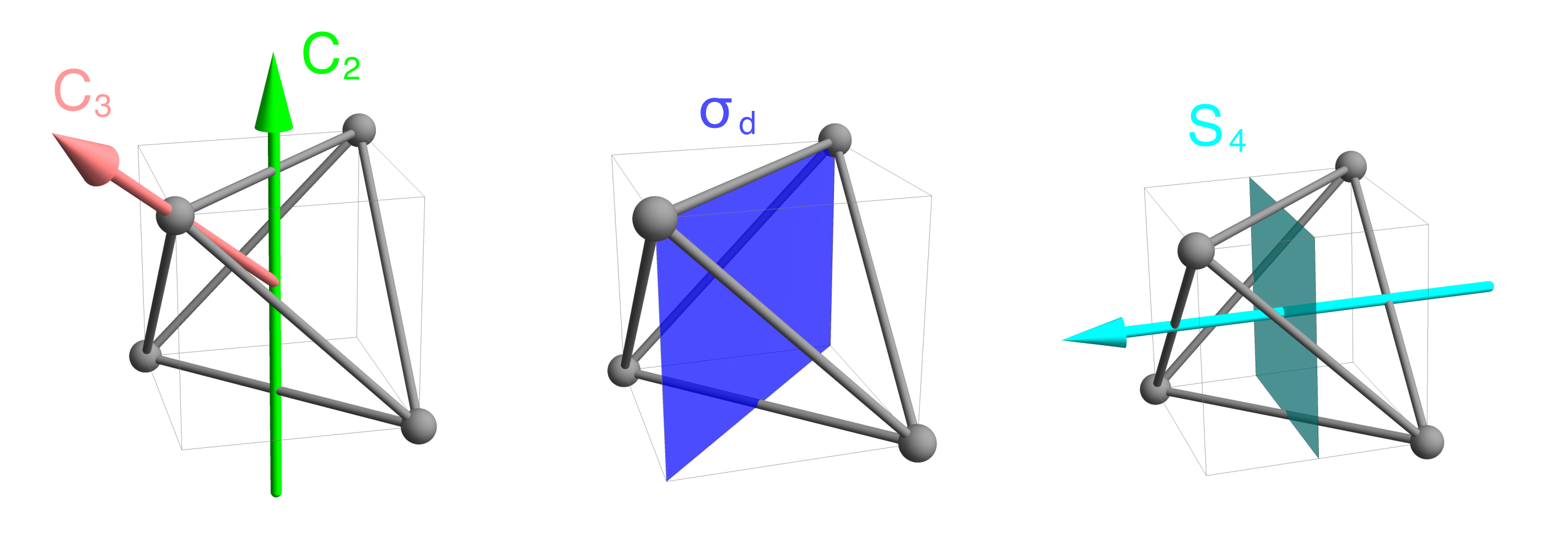}
\caption{Point group symmetries of the pyrochlore lattice. The arrows and planes represent the rotation axes and reflection planes of the point group symmetries $\{C_3,C_2,\sigma_d,S_4\}$.}
\label{fig:symm}
\end{figure}

Space group symmetries of the pyrochlore iridates (number 227; $Fd\bar{3}m$) play crucial roles in the symmetry protection of the triple-point and nodal-line topological magnons discussed in the main text.
$T_d$ point group is a subgroup of $Fd\bar{3}m$ containing symmetries relevant to the triply degenerate band crossings.
\begin{equation}
T_d=\textup{Span}\{ C_3, C_2, \sigma_d, S_4\}.
\end{equation}
The four generators are threefold rotation about $\langle$111$\rangle$ axis ($C_3$), twofold rotation about $\langle$100$\rangle$ axis ($C_2$), mirror reflection with respect to $\{$110$\}$ plane ($\sigma_d$), and $90^{\circ}$ rotation about $\langle$100$\rangle$ axis followed by reflection ($S_4$);
see Fig. \ref{fig:symm}.

\begin{table*}
\begin{ruledtabular}
\caption{Symmetry operations in the AIAO state. The second column defines each operation in the coordinate system of Fig. \ref{fig:structure}. The associated representation matrices in magnon basis are provided in the third column.}
\label{tab:symm}
\begin{tabular}{lcc}
$\mathcal{O}$ (symmetry operation) & $\mathcal{O}: {\bf r}=(x,y,z)\rightarrow{\bf r}'=(x',y',z')$ & $U_\mathcal{O}$
\\
\hline
$C_3$ [rotation axis: $(x,x,x)$] & $(y,z,x)$ &
$\left( 
\begin{array}{cccc}
 e^{i2\pi/3}&0&0&0
 \\
 0&0&e^{i\pi/3}&0
 \\
 0&0&0&-1
 \\
 0&e^{i2\pi/3}&0&0
\end{array}
\right)$
\\
$C_2$ [rotation axis: $(1/2,y,1/2)$] & $(-x+1,y,-z+1)$ &
$\left( 
\begin{array}{cccc}
 0&0&1&0
 \\
 0&0&0&1
 \\
 1&0&0&0
 \\
 0&1&0&0
\end{array}
\right)$
\\
\hline
$\bar{S}_4$ 
$\left[
\begin{array}{l}
\textup{rotation~axis}:~(x,1/2,1/2)
\\
\textup{reflection~plane}:~(1/2,y,z)
\end{array}
\right]$
 & $(-x+1,-z+1,y)$ &
$\left( 
\begin{array}{cccc}
 0&0&e^{i\pi/3}&0
 \\
 0&0&0&-1
 \\
 0&e^{i2\pi/3}&0&0
 \\
 e^{i2\pi/3}&0&0&0
\end{array}
\right)$
\\
\hline
$G_d$
$\left[
\begin{array}{l}
\textup{reflection plane}:~(x,y,0)
\\
\textup{translation by}~(1,1,0)
\end{array}
\right]$
& $(x+1,y+1,-z)$ &
$\left( 
\begin{array}{cccc}
 0&0&0&e^{i\pi/3}
 \\
 0&0&e^{-i\pi/3}&0
 \\
 0&e^{i\pi/3}&0&0
 \\
 e^{-i\pi/3}&0&0&0
\end{array}
\right)$
\end{tabular}
\end{ruledtabular}
\end{table*}

In the presence of the AIAO order, the two improper rotations ($\sigma_d,S_4$) are no longer symmetries of the system (note that spin is invariant under spatial inversion).
Instead, the AIAO state is invariant under the improper rotations multiplied by time reversal ($\Theta$), leading to the following magnetic point group.
\begin{equation}
\bar{T}_d=\textup{Span}\{ C_3, C_2, \bar{\sigma}_d, \bar{S}_4\} 
\end{equation}
where $\bar{\mathcal{O}}=\Theta * \mathcal{O}$ for $\mathcal{O}$=$\sigma_d$, $S_4$.
These magnetic point group symmetries protect the triply degenerate crossings (TDC) in the magnon bands of the AIAO state.

Specifically, the A-type TDC at $\Gamma$ of the Brillouin zone is a three-dimensional irreducible representation of the subgroup $T=\textup{Span}\{C_3, C_2\}$.
To explicitly show this, we consider 
the $C_3$ rotation about a $[\bar{1}\bar{1}\bar{1}]$ axis and the $C_2$ rotation about a [010] axis defined in Table \ref{tab:symm}.
\begin{eqnarray}
C_3: \Psi_{\bf k}
&\rightarrow&
\left(
\begin{array}{cc}
U_{C_3} & 0
\\
0 & U_{C_3}^*
\end{array}
\right)
\Psi_{(k_z,k_x,k_y)}
\\
C_2: \Psi_{\bf k}
&\rightarrow&
e^{-i(k_x+k_z)}
\left(
\begin{array}{cc}
U_{C_2} & 0
\\
0 & U_{C_2}^*
\end{array}
\right)
\Psi_{(-k_x,k_y,-k_z)}~~~~~
\label{eq:C2}
\end{eqnarray}
The above equations show the transformation rules of magnon operators under the $C_3$ and $C_2$ rotations. The $4\times4$ representation matrices, $U_{C_3}$ and $U_{C_2}$, are listed in Table \ref{tab:symm}. 
The eigenvalues of the matrices are given by $\{1,e^{i2\pi/3},e^{-i2\pi/3},e^{i2\pi/3} \}$ for $U_{C_3}$ and $\{ +1,+1,-1,-1 \}$ for $U_{C_2}$.
From this eigenvalue structure, one can easily check that four magnon energy levels at the $\Gamma$ point comprise three-dimensional representation and one-dimensional representation of the $T$ point group.

The B-type TDC is a crossing between nondegenerate and doubly degenerate bands along $\Gamma$X of the Brillouin zone. The twofold degeneracy of the latter is provided by the $\bar{S_4}~(=\Theta * S_4)$ anti-unitary symmetry defined in Table \ref{tab:symm}.
The corresponding magnon transformation rule is given by
\begin{equation}
\bar{S}_4: \Psi_{\bf k}\rightarrow
e^{-i(k_x+k_z)}
\left(
\begin{array}{cc}
U_{\bar{S}_4} & 0
\\
0 & U_{\bar{S}_4}^*
\end{array}
\right)
\mathcal{K}
\Psi_{(k_x,k_z,-k_y)},
\end{equation}
where $\mathcal{K}$ means complex conjugation originating from time reversal, and the $4\times4$ matrix $U_{\bar{S}_4}$ is shown in Table \ref{tab:symm}.
We can show that along $\Gamma\textup{X}=(k_x,0,0)$ the representation matrix of $(\bar{S_4})^2$, which is unitary and given by $U_{\bar{S}_4} U_{\bar{S}_4}^*$ for the  particle sector, has the eigenvalues $\{+1,+1,-1,-1\}$.
The doubly degenerate band along $\Gamma$X has the eigenvalue $(\bar{S_4})^2=-1$, which guarantees at least twofold degeneracy in a similar way to the Kramer's degeneracy.

\begin{figure}
\centering
\includegraphics[width=0.75\linewidth]{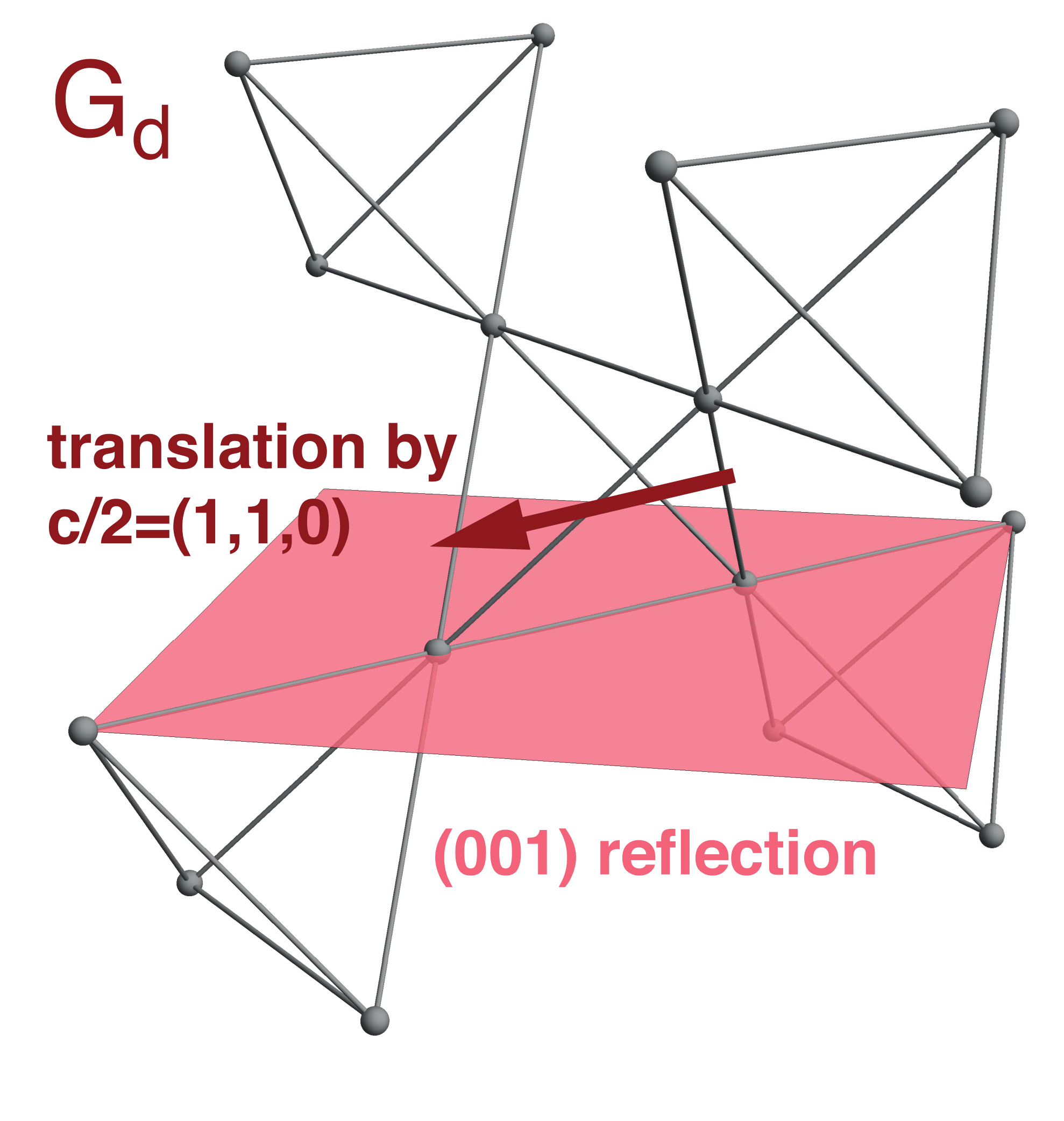}
\caption{Nonsymmorphic glide symmetry $G_d$.}
\label{fig:glide}
\end{figure}

Nonsymmorphic glide $G_d$ (an element of $Fd\bar{3}m$) is an important symmetry that protects the nodal line band crossings along XW of the Brillouin zone.
The glide symmetry is a mirror reflection about a (001) plane combined with a fractional translation by ${\bf c}/2=(1,1,0)$, as illustrated in Fig. \ref{fig:glide} and Table \ref{tab:symm}.
Since $(G_d)^2=\mathcal{T}_{\bf c}$ (translation by ${\bf c}$), the allowed eigenvalues of the glide symmetry are $\pm e^{-i{\bf k} \cdot {\bf c}/2}$.
Interesting, unlike the improper rotations $\sigma_d$ and $S_4$, the glide $G_d$ itself is a symmetry of the AIAO state without combination with time reversal. 
Acting on magnon operators, the glide symmetry has the following representation:
\begin{equation}
G_d: \Psi_{\bf k}\rightarrow
e^{-i(k_x+k_y)}
\left(
\begin{array}{cc}
U_{G_d} & 0
\\
0 & U_{G_d}^*
\end{array}
\right)
\Psi_{(k_x,k_y,-k_z)}
\label{eq:Gd}
\end{equation}
where the matrix $U_{G_d}$ is shown in Table \ref{tab:symm}.
Along $\textup{XW}=(\pi/2,k_y,0)$, the little group at each ${\bf k}$ point includes the $G_d$ as well as the aforementioned $C_2$ about the [010] axis (Table \ref{tab:symm}). Their representation matrices have the eigenvalues $\{+e^{-i{\bf k} \cdot {\bf c}/2},+e^{-i{\bf k} \cdot {\bf c}/2},-e^{-i{\bf k} \cdot {\bf c}/2},-e^{-i{\bf k} \cdot {\bf c}/2}\}$ for $G_d$ and $\{+1,+1,-1,-1\}$ for $C_2$. The two little group elements anticommute with each other as can be verified from Eqs. \ref{eq:C2} and \ref{eq:Gd}.
Hence, $| G_d= + e^{-i{\bf k} \cdot {\bf c}/2} \rangle  \xrightleftharpoons[]{C_2} | G_d= - e^{-i{\bf k} \cdot {\bf c}/2} \rangle$ and $| C_2= + 1 \rangle  \xrightleftharpoons[]{G_d} | C_2= - 1 \rangle$.
This anticommuting nature of $G_d$ and $C_2$ along XW ultimately leads to the double degeneracy at each energy level along the ${\bf k}$ line.

\begin{figure}[b]
\centering
\includegraphics[width=\linewidth]{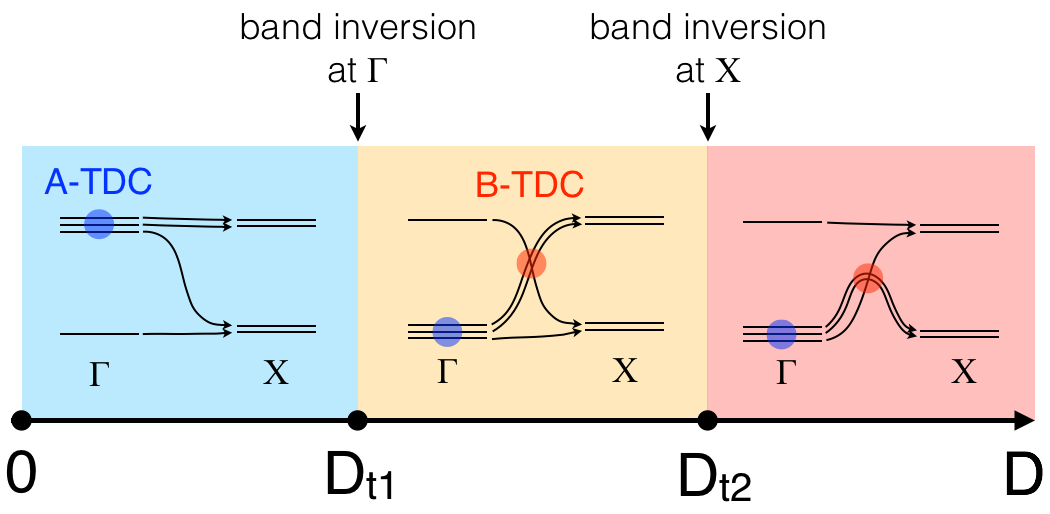}
\caption{Schematic illustration of the topological transitions in the magnon band structure.}
\label{fig:schematic-topological-transition}
\end{figure}

\section{Magnon band structure}

\subsection{Zero field}

\begin{figure}[b]
\centering
\includegraphics[width=\linewidth]{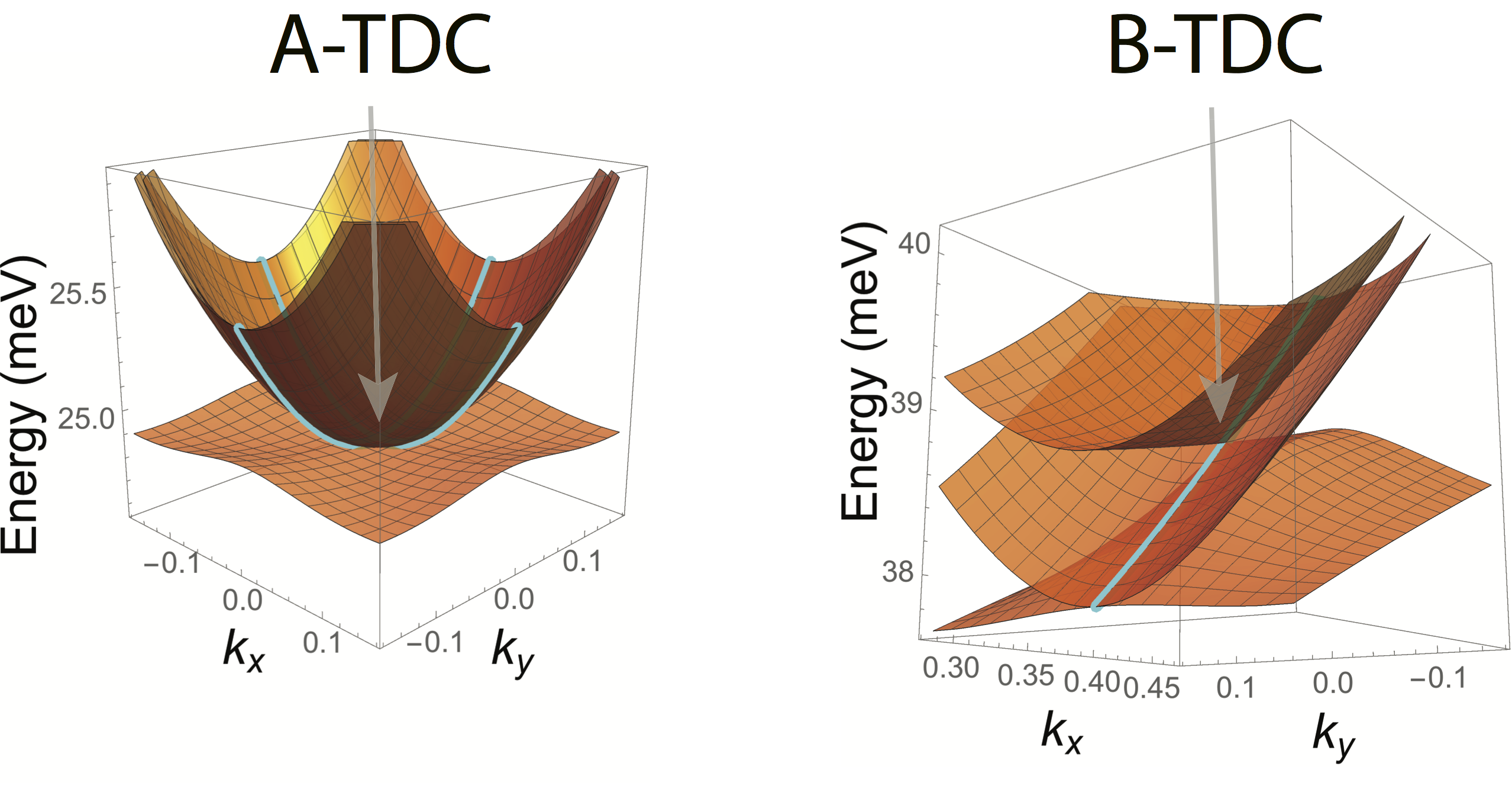}
\caption{Triply degenerate crossings of the magnon bands. Left: A-type TDC at the $\Gamma$ point ($D=0.18J$). Right: B-type TDC occurring along the $\Gamma$X line ($D=0.33J$).
In both cases, $k_z=0$, and the cyan curves indicate the doubly degenerate energy levels protected by $\Theta * S_4$.}
\label{fig:TDC}
\end{figure}

\begin{figure}
\centering
\includegraphics[width=\linewidth]{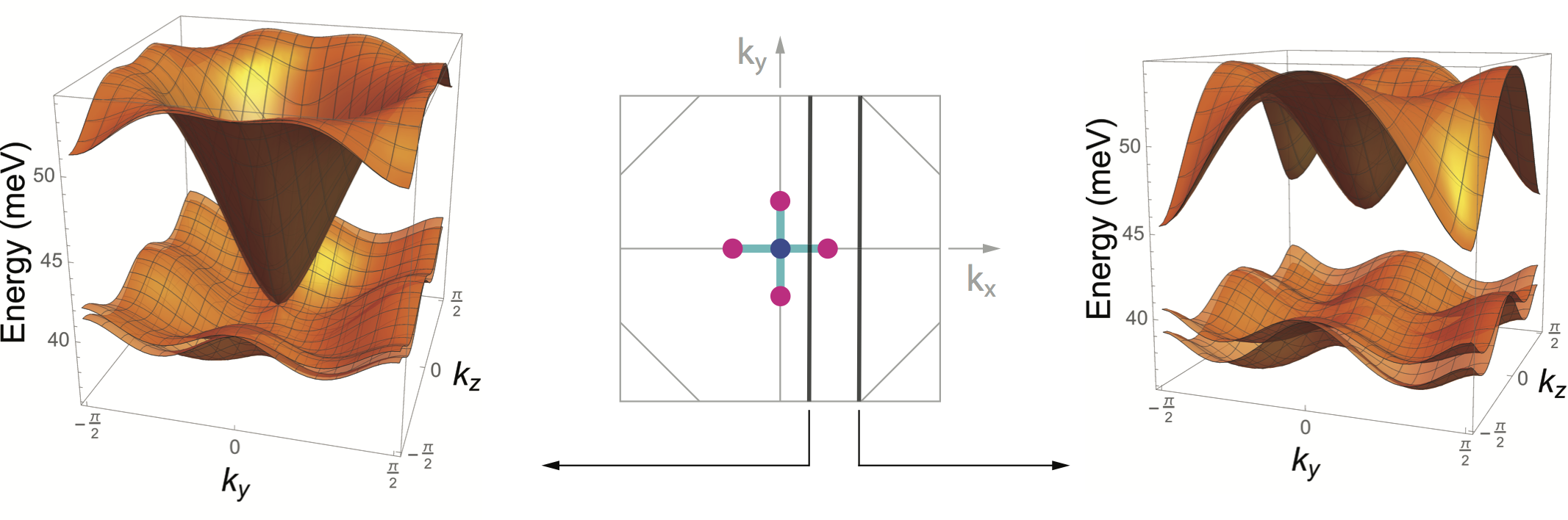}
\caption{Magnon band structures on two different planes in the Brillouin zone for $D=0.33J$.
Middle: The configuration of the triple points (blue: A-type, red: B-type) on the $k_z=0$ plane of the Brillouin zone. The cyan lines indicate the ${\bf k}$-points where the doubly degenerate bands (protected by $\Theta * S_4$) appear.
Left: ``Metallic" band structure on a $k_y$-$k_z$ plane with $k_x<k_x^\textup{B-TP}$.
Right: ``Insulating" band structure on a $k_y$-$k_z$ plane with $k_x>k_x^\textup{B-TP}$.
Here $k_x^\textup{B-TP}$ means the $x$ coordinate of the B-type triple point sandwiched by the two black lines in the middle figure.}
\label{fig:TP}
\end{figure}

As discussed in the main text, the magnon band structure undergoes topological transitions at $D=D_{t1}$ and $D=D_{t2}$. 
The transitions occur through band inversions at the $\Gamma$ and X points as illustrated in Fig. \ref{fig:schematic-topological-transition}.
The energy levels at the $\Gamma$ and X points are given by
\begin{eqnarray}
\Gamma:
~~~
&&
E_1({\Gamma})=2S\sqrt{\frac{8\sqrt{2}JD+14D^2}{3}}~\textup{(triplet)}, 
\nonumber\\
&&
E_2({\Gamma})=2S \cdot 3\sqrt{2}D~\textup{(singlet)},
\end{eqnarray}
\begin{eqnarray}
\textup{X}:
~~~
&&
E_1({\textup{X}})=2S\sqrt{\frac{8\sqrt{2}JD+32D^2}{3}}~\textup{(doublet)},
\nonumber\\
&&
E_2({\textup{X}})=2S\sqrt{\frac{4J^2+12\sqrt{2}JD+16D^2}{3}}~\textup{(doublet)}.
\nonumber\\
\end{eqnarray}
It is straightforward to check that $D_{t1}=J\sqrt{2}/5~(=0.28J)$ from the equation $E_1({\Gamma})=E_2({\Gamma})$, and $D_{t2}=J/\sqrt{2}~(=0.71J)$ from the equation $E_1(X)=E_2(X)$.
The magnon band structure of $H_{\textup{SW}}$ has an energy gap given by $\Delta\equiv\textup{min}\{E_1({\Gamma}),E_2({\Gamma})\}$.

Figure \ref{fig:TDC} visualizes triply degenerate band crossings (TDC) of the magnon bands.
In the A-type TDC, three quadratically dispersing bands are touching at the $\Gamma$ point (Fig. \ref{fig:TDC} left).
On the other hand, the B-type TDC is understood as a type-II Weyl node with an extra linearly dispersing band that is touching the Weyl cone (Fig. \ref{fig:TDC} right).
Despite such similarity, there is a crucial difference between the B-type triple points and Weyl points.
In the Brillouin zone, a Weyl point can be considered as a topological transition point between two-dimensional trivial insulator and Chern insulator.
Both sides of the Weyl point is ``insulating" with a finite energy gap.
However, the B-type triple point is viewed as a transition point between two-dimensional metal and insulator (see Fig. \ref{fig:TP}). Here, insulator (metal) means that the system has a finite (zero) gap between the top two and bottom two bands.

\begin{figure}
\centering
\includegraphics[width=0.9\linewidth]{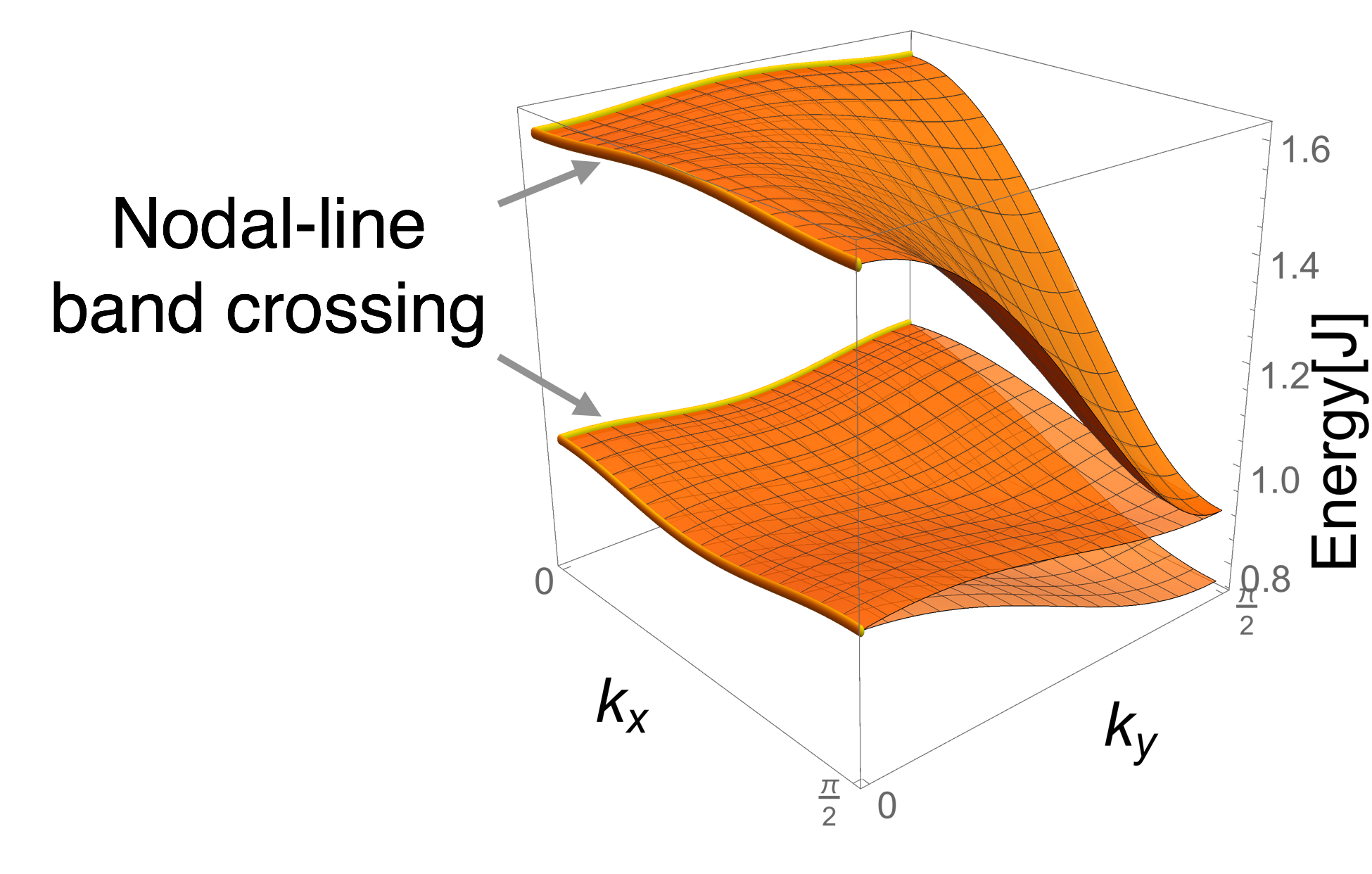}
\caption{Doubly degenerate nodal-line band crossings.}
\label{fig:NLC}
\end{figure}

Figure \ref{fig:NLC} illustrates nodal-line band crossings on the $k_z=\pi/2$ plane of the Brillouin zone. Along the $k_x=0$ line and $k_y=0$ line (symmetry related lines of XW), each pair of the upper two and lower two bands become degenerate due to the anticommuting nature of $C_3$ and $C_2$ rotations along the ${\bf k}$ line. On the other hand, the doubly degenerate band occurring along the $k_x=\pi/2$ line and $k_y=\pi/2$ (symmetry related lines of $\Gamma$X) is also nodal line band crossing. In this case, twofold degeneracy occurs only in the top two bands which have the eigenvalue $(\bar{S}_4)^2=-1$.

\subsection{Nonzero field}

\begin{figure}[b]
\vspace{-0.5cm}
\centering
\includegraphics[width=\linewidth]{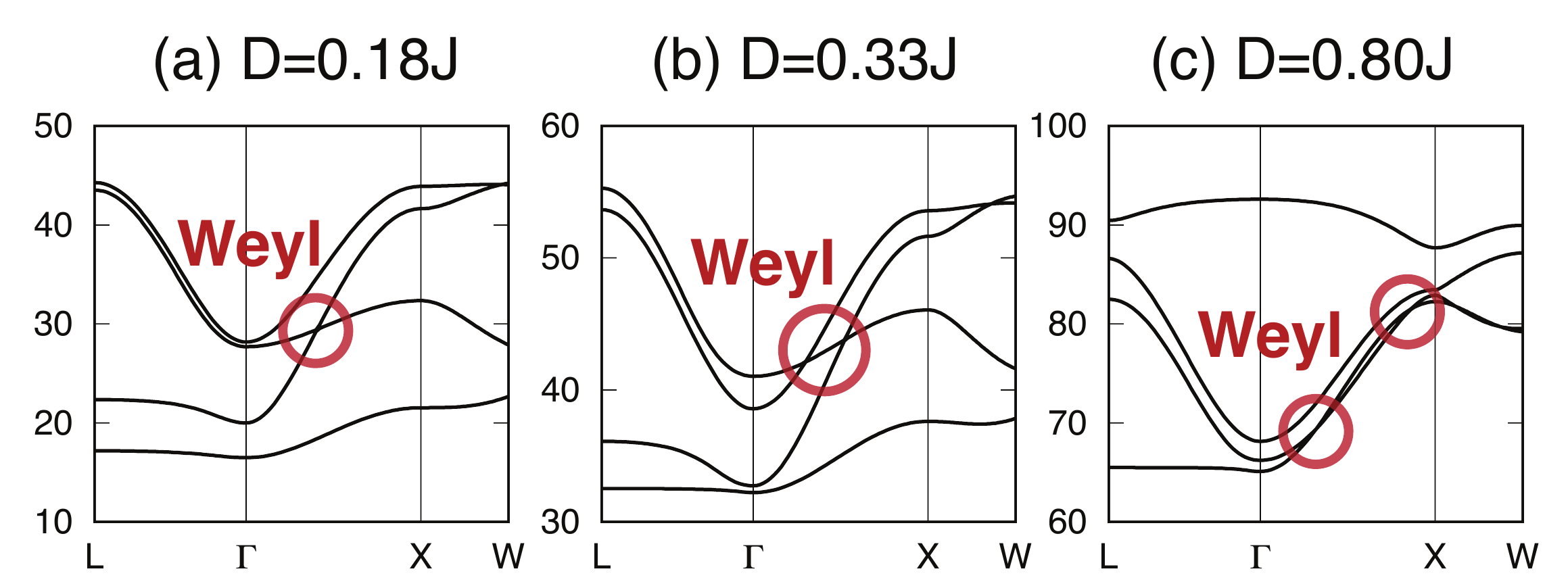}
\caption{
Magnon band structures under a magnetic field $h=0.3J//[110]$ for (a) $D=0.18J$, (b) $D=0.33J$, and (c) $D=0.80J$.
The Weyl points created by the field are highlighted by circles.
In each plot, the vertical axis represents energy in the unit of meV.
\label{fig:magnon_bands}
}
\end{figure}

Under a magnetic field, the triple points are no longer protected since associated symmetries are broken by the field.
Instead, type-II Weyl points \cite{Soluyanov2015} are created from the A- and B-TDC as shown in Fig. \ref{fig:magnon_bands} (compare with Fig. 2).
For purposes of clarity we have chosen a large magnetic field ($h=0.3J\slantedparallel[100]$), though the Weyl points appear for arbitrary field strength and direction; the distance between a pair of Weyl points increases with field.

\begin{figure}[b]
\vspace{-0.5cm}
\centering
\includegraphics[width=\linewidth]{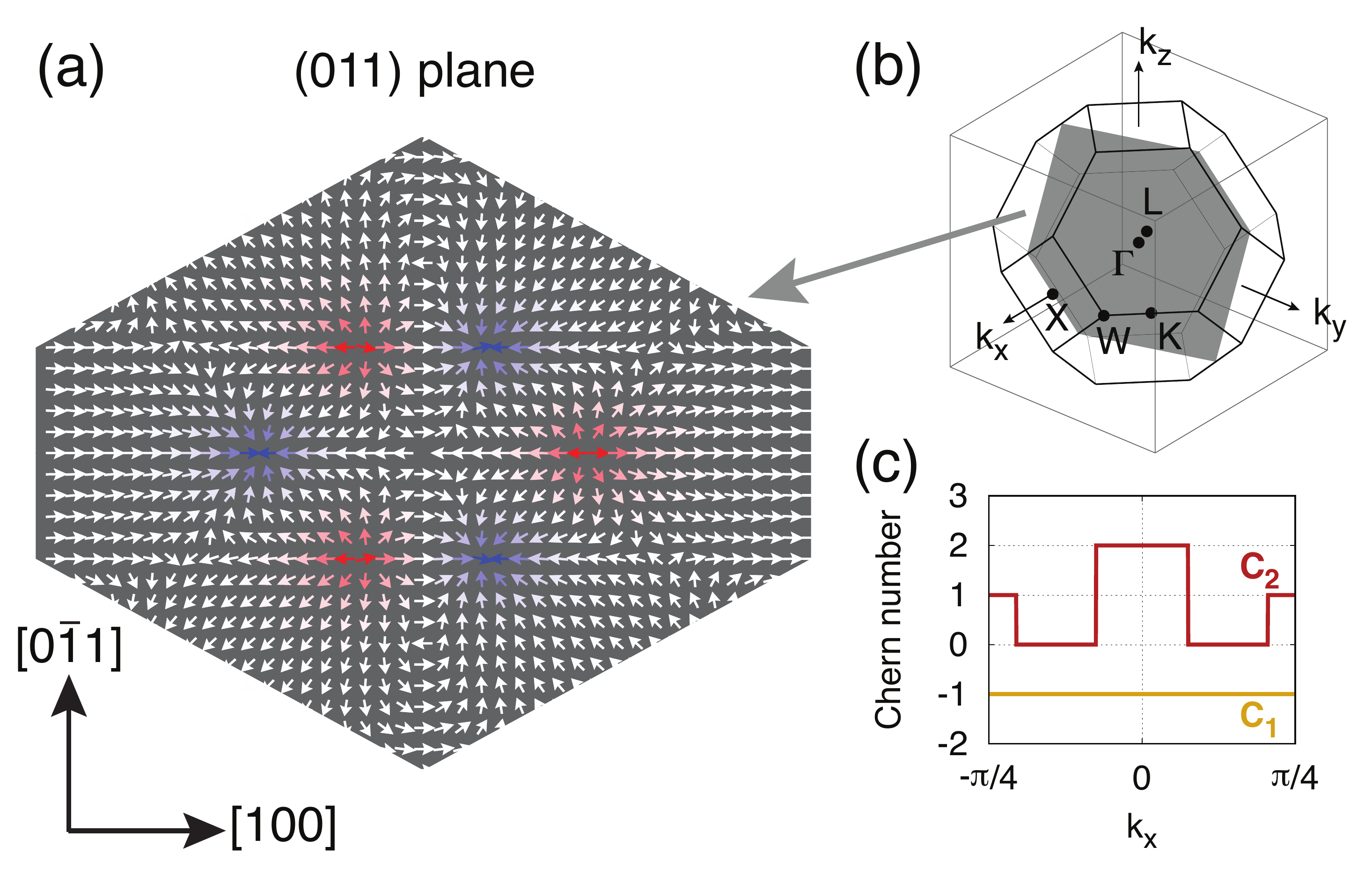}
\caption{
(a) Normalized Berry curvature $\hat{\boldsymbol \Omega}_{2{\bf k}}$ of the second lowest band in Fig. \ref{fig:magnon_bands} (a). The red and blue highlights the Weyl points appearing on the (110) plane.
(b) Brillouin zone with several high symmetry points [$\Gamma:(0,0,0)$, $\textup{X}:(\pi/2,0,0)$, $\textup{W}:(\pi/2,\pi/4,0)$, $\textup{K}:(3\pi/8,3\pi/8,0)$, $\textup{L}:(\pi/4,\pi/4,\pi/4)$].
(c) Chern numbers $C_1$ and $C_2$ for the lowest and second lowest bands in Fig. \ref{fig:magnon_bands} (a).
}
\label{fig:Weyl_Berry}
\end{figure}

The existence of Weyl points also manifests as sources and sinks for the Berry curvature of the magnon bands.
For each magnon band $(n=1,\cdots,4)$, the Berry curvature $\boldsymbol{\Omega}_{n{\bf k}}=({\Omega}_{n{\bf k}}^x,{\Omega}_{n{\bf k}}^y,{\Omega}_{n{\bf k}}^z)$ is defined in terms of the Bogoliubov transformation matrix $T_{\bf k}$ \cite{Matsumoto2014}:
\begin{equation}
\boldsymbol{\Omega}_{n{\bf k}} = i \sum_{m=1}^8 \frac{\partial [T_{{\bf k}}^{-1}]_{nm}}{\partial {\bf k}} \boldsymbol{\times}  \frac{\partial [T_{{\bf k}}]_{mn}}{\partial {\bf k}}.
\label{eq:Berry}
\end{equation}
Figure \ref{fig:Weyl_Berry} (a) shows the Berry curvature $\boldsymbol{\Omega}_{2{\bf k}}$ for the second lowest band in Fig. \ref{fig:magnon_bands} (a) ($D=0.18J,~h=0.3J\slantedparallel[100]$) as an example.
In this case, we find three pairs of Weyl points at the (011) plane passing through the $\Gamma$ point. 
The Weyl points are highlighted by red (source) and blue (sink).
Their topological charges (+1/$-$1 for a source/sink Weyl point) are confirmed by the change in the Chern number $C_2(k_x)\equiv\frac{1}{2\pi}\int_{-\pi}^{\pi} dk_y \int_{-\pi}^{\pi} dk_z {\Omega}_{2{\bf k}}^x$ [Fig. \ref{fig:Weyl_Berry} (c)].
Such Weyl magnon excitations have been also found in other pyrochlore magnets \cite{Li2016,Mook2016}.

\section{Thermal Hall effect}

\begin{figure}
\centering
\includegraphics[width=0.45\linewidth]{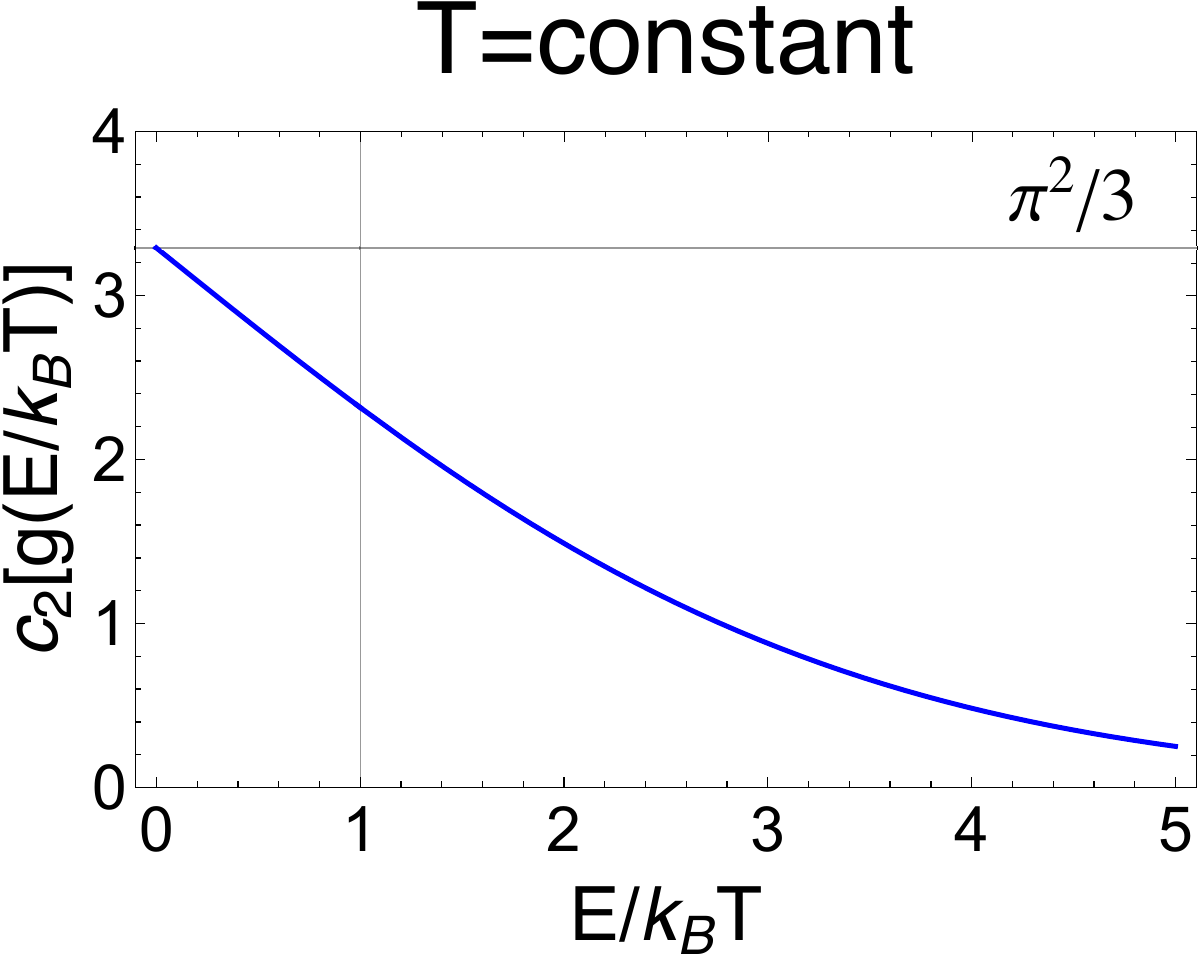}
~~~
\includegraphics[width=0.45\linewidth]{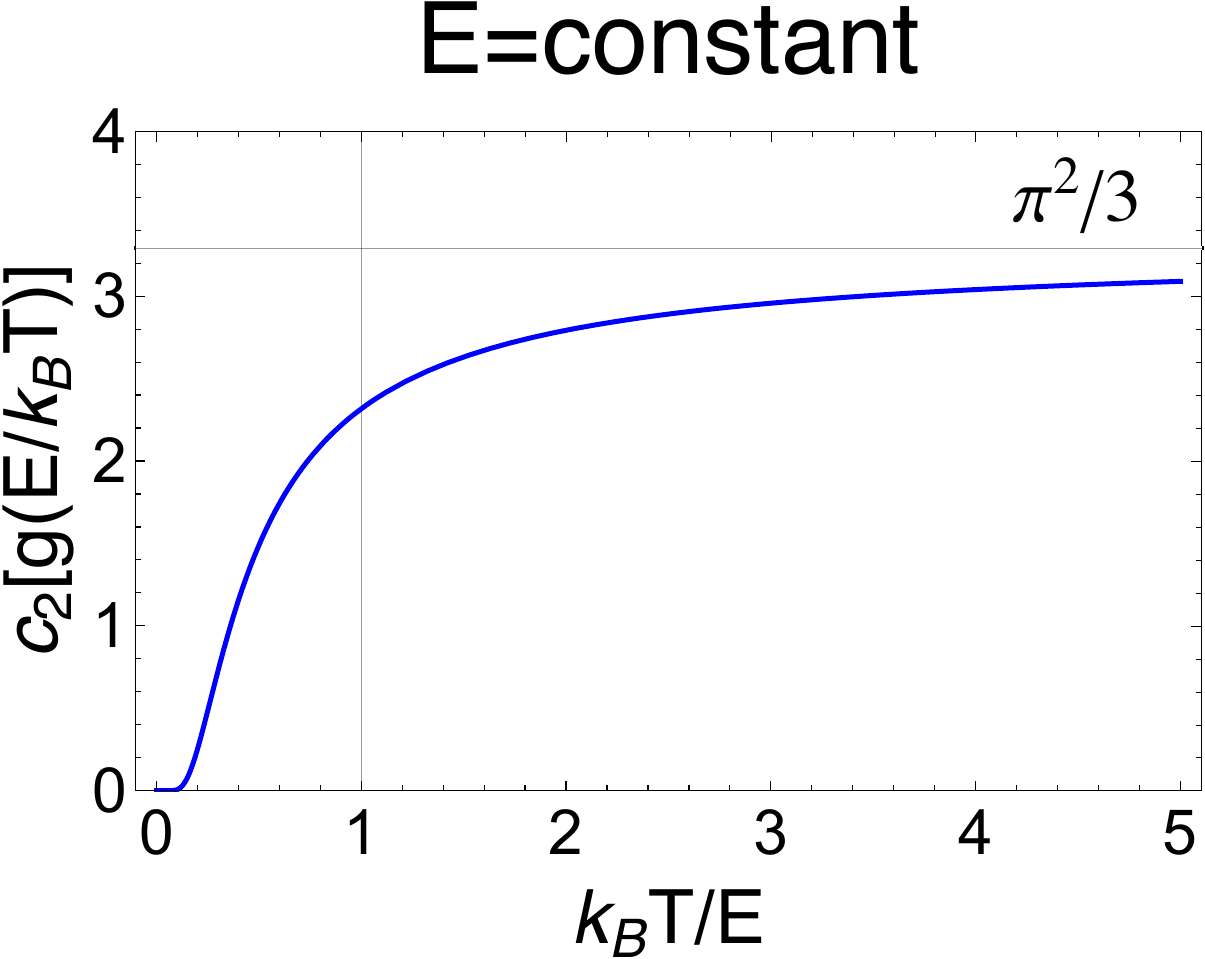}
\caption{Energy and temperature dependence of the weight function $c_2[g(E/k_BT)]$.}
\label{fig:c2-ftn}
\end{figure}

As shown in the Kubo formula [Eq. 4], magnon thermal Hall conductivity is determined by the two factors, magnon Berry curvature $\Omega_{n{\bf k}}^{\rho}$ and weight function $c_2[g(E_{n{\bf k}}/k_BT)]$.
The Berry curvature can be expressed in terms of the velocity operator ${\bf v}_{\bf k}~(\equiv \nabla_{\bf k} H_{\bf k})$ \cite{Matsumoto2014}:
\begin{equation}
\Omega_{n{\bf k}}^{\rho} = i \epsilon_{\rho\mu\nu} \sum_{m(\ne n)=1}^8 \frac{ (\sigma_3 \bar{v}_{\bf k}^{\mu})_{nm} (\sigma_3 \bar{v}_{\bf k}^{\nu})_{mn} }{[(\sigma_3E_{\bf k})_n-(\sigma_3E_{\bf k})_m]^2},
\end{equation}
where $\bar{\bf v}_{\bf k}=T_{\bf k}^{\dagger} {\bf v}_{\bf k} T_{\bf k}$.
Equivalence of this expression to Eq. 3 can be checked using the identity
$
\bar{\bf v}_{\bf k}
=
\frac{\partial E_{\bf k}}{\partial {\bf k}} 
- 
[\sigma_3 E_{\bf k}, T_{\bf k}^{\dagger} \sigma_3 \frac{\partial T_{\bf k}}{\partial {\bf k}}] .
$
The weight function $c_2[g(E/k_BT)]$ is semi-positive and has the following two limits.
\begin{equation}
c_2[g(E/k_BT)]=
\left\{
\begin{array}{cl}
\pi^2/3 & (E/k_BT=0)
\\
0 & (E/k_BT\rightarrow\infty)
\end{array}
\right\} .
\end{equation}
The full temperature and energy dependence of the function is illustrated in Fig. \ref{fig:c2-ftn}.

\begin{figure}
\centering
\includegraphics[width=\linewidth]{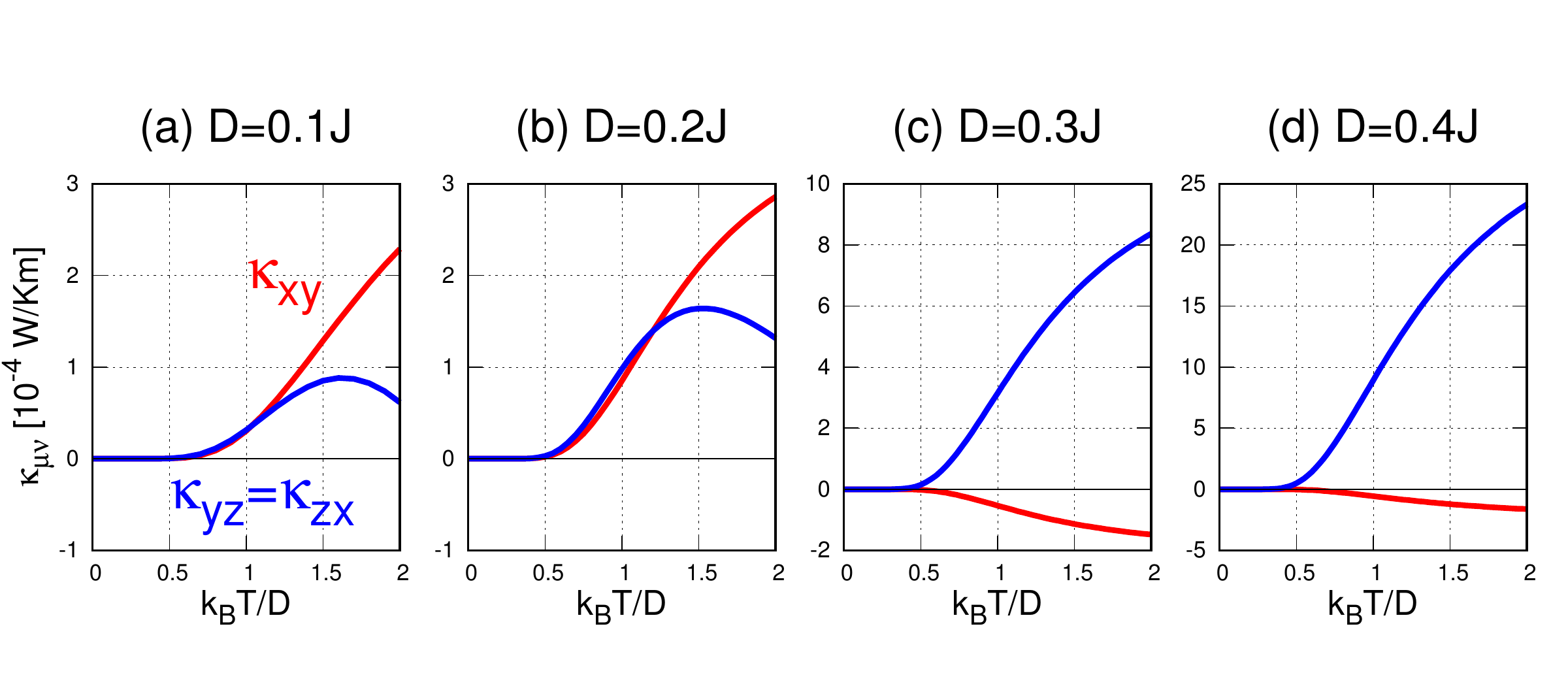}
\caption{Thermal Hall conductivity for $D=0.1J,~0.2J,~0.3J,~0.4J$, under a magnetic field $h=0.02J//[110]$.}
\label{fig:kappa_T_h110_additional}
\end{figure}

In Fig. 4 of the main text, we have shown that the system exhibits distinct patterns of thermal Hall conductivity in the regimes I and II using two exemplary cases ($D=0.18J,~0.33J$). In Fig. \ref{fig:kappa_T_h110_additional}, we present further calculation results obtained for other values of the DM interaction ($D=0.1J,~0.2J,~0.3J,~0.4J$) in the presence of a small magnetic field $h=0.02J$ along the [110] direction.
One can clearly see that the characteristic thermal Hall responses shown in Fig. 4 are generic features of the two band topology regimes from Fig. \ref{fig:kappa_T_h110_additional}.

\begin{figure}
\centering
\includegraphics[width=\linewidth]{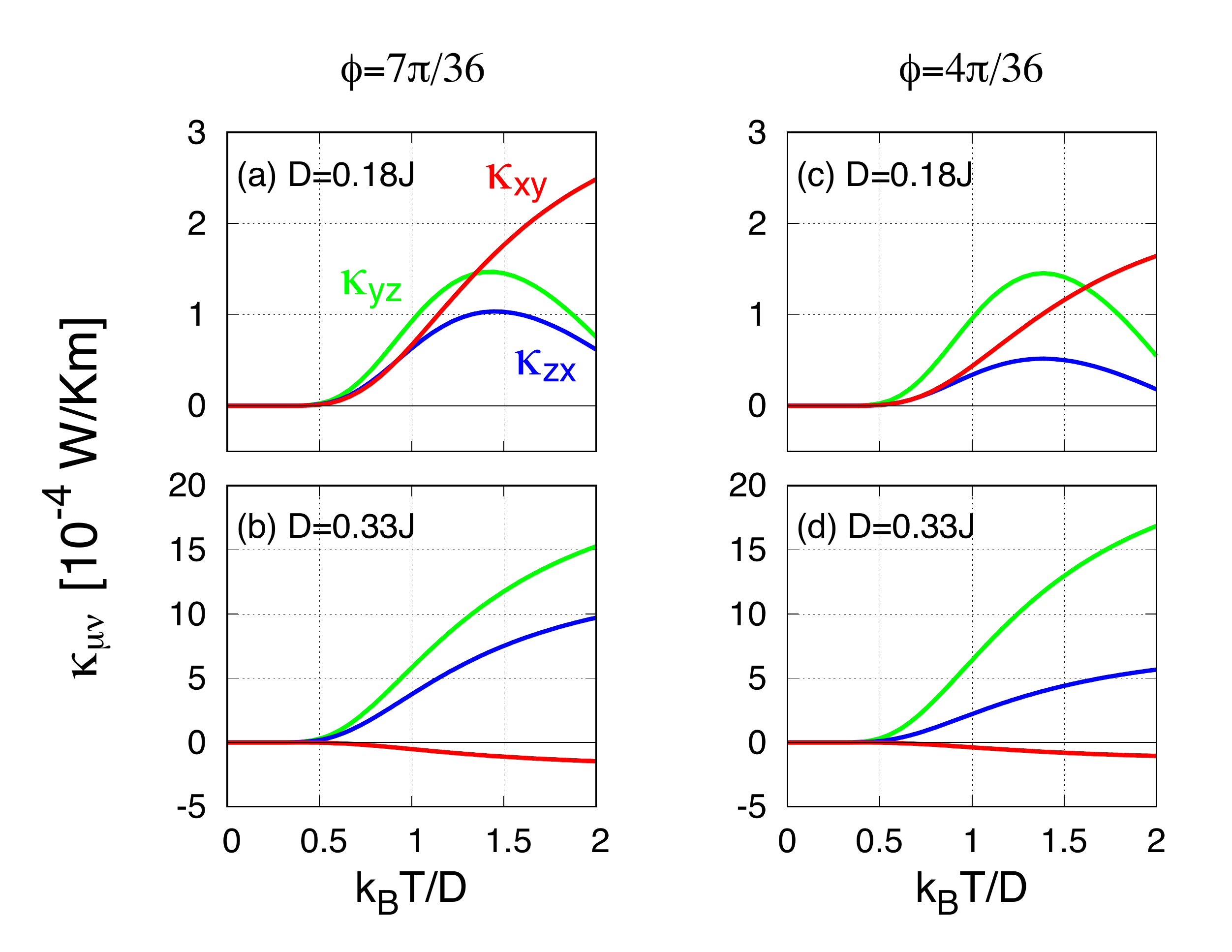}
\caption{Thermal Hall conductivity for field directions between the [110] and [100] axes.
(a,b) $\phi={7\pi}/{36}$ and (c,d) $\phi={4\pi}/{36}$ with $\hat{h}=\hat{x}\textup{cos}\phi  + \hat{y}\textup{sin}\phi$. In each case, $h=0.02J$.}
\label{fig:kappa_T_other_directions}
\end{figure}

Remarkably, such characteristic behaviors are far more generic, not restricted to the [110] field direction.
Figure \ref{fig:kappa_T_other_directions} shows the thermal Hall conductivity for two different field directions between [110] and [100]: $\phi={7\pi}/{36}$ (left) and $\phi={4\pi}/{36}$ (right) with $\hat{h}= \hat{x}\textup{cos}\phi  + \hat{y}\textup{sin}\phi$.
Due to the field direction deviation from the [110] axis, the two components $\kappa_{yz}$ (green) and $\kappa_{zx}$ (blue) are not any longer identical to each other.
Nevertheless, we find qualitatively same behaviors as shown in Figs. 3 (c,d).

\section{Extended model with further neighbor interactions}

\begin{figure}
\centering
\includegraphics[width=\linewidth]{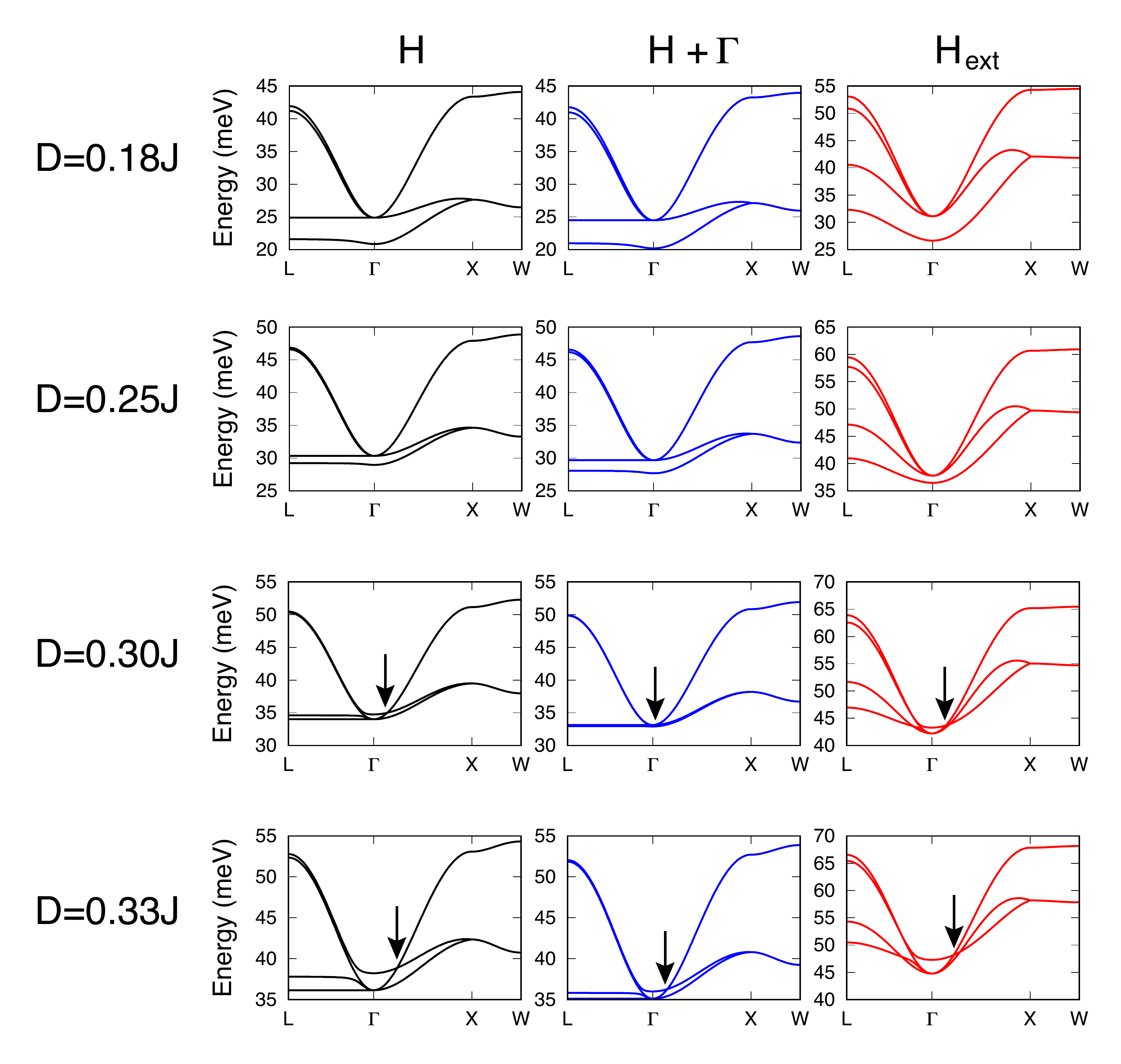}
\caption{Magnon band structures of three models. Left: the original $J$-$D$ model $H$. Middle: $H$ supplemented with the nearest-neighbor $\Gamma$ interaction. Right: the extended model $H_{\textup{ext}}$ with $J'=0.1J$, $D'_1=-0.1D$, $D'_2=-0.1D$. The B-type TDC is marked by an arrow. In the case of $H$ supplemented with the NN $\Gamma$ interaction at $D=0.30J$, the B-type TDC occurs very close to the $\Gamma$ point of the Brillouin zone.}
\label{fig:magnon-bands}
\end{figure}

\begin{figure}
\centering
\includegraphics[width=0.8\linewidth]{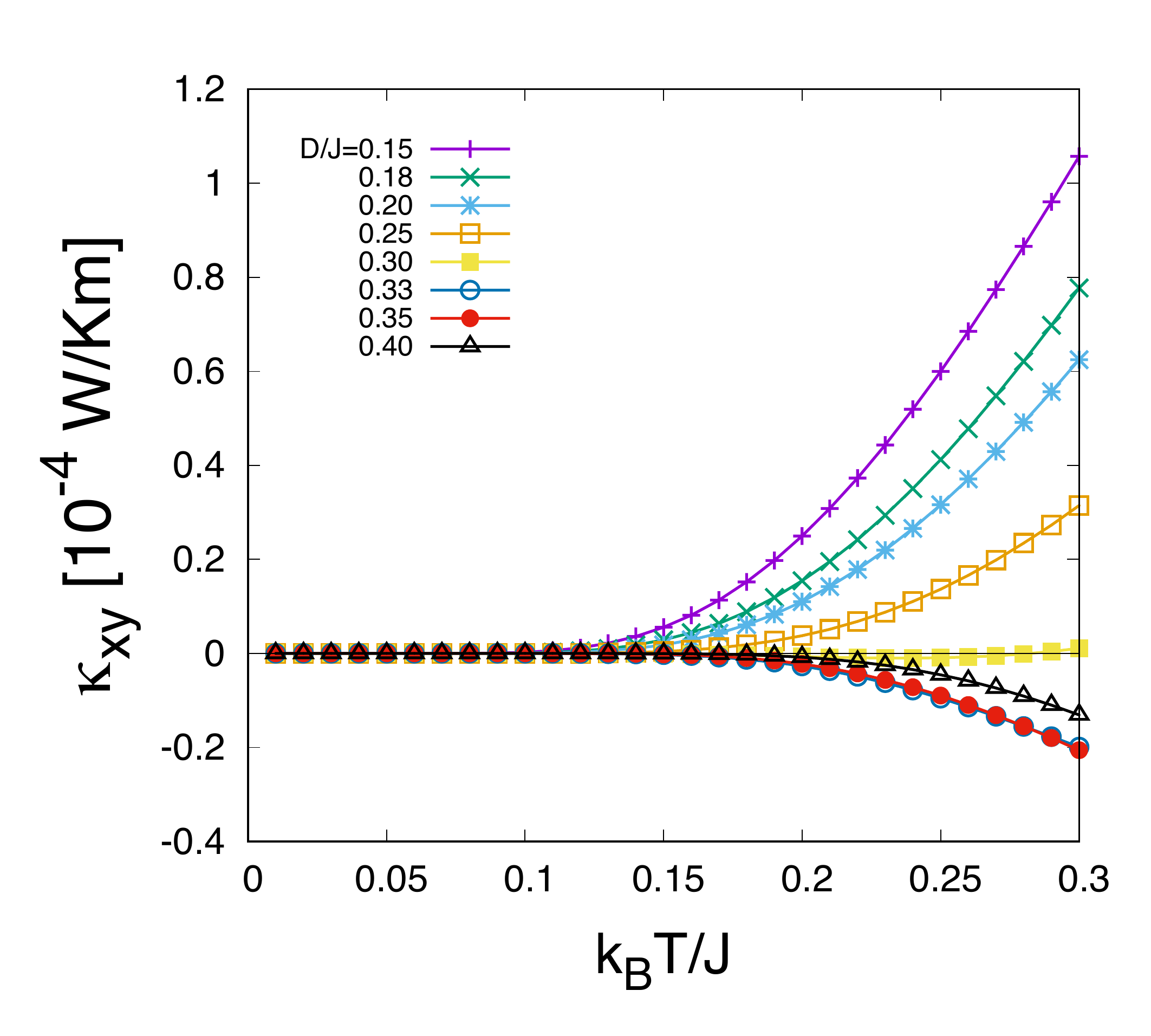}
\caption{$\kappa_{xy}$ of the extended model $H_{\textup{ext}}$ with $J'=0.1J$, $D'_1=-0.1D$, $D'_2=-0.1D$ under an external field $h=0.02J$//[110].}
\label{fig:kappa-xy-NNNmodel}
\end{figure}

Here we examine the robustness of the TDC and characteristic thermal Hall responses by extending our model to second nearest-neighbor interactions allowed by symmetry.
In the extended model $H_{\textup{ext}}$, the original nearest-neighbor $J$-$D$ model $H$ is perturbed by second-nearest-neighbor $J'$, ${\bf D}'_{ij}$, and $\Gamma'_{ij}$ terms as well as nearest-neighbor $\Gamma$ term.
\begin{eqnarray}
H_{\textup{ext}}
&=&
H
+\sum_{ij \in \textup{NN}} \Gamma_{ij}^{\mu\nu} S_i^{\mu} S_j^{\nu}
\nonumber\\
&+&
\sum_{ij \in \textup{NNN}} J' {\bf S}_i \cdot {\bf S}_j + {\bf D}'_{ij} \cdot {\bf S}_i \times {\bf S}_j + {\Gamma'_{ij}}^{\mu\nu} S_i^{\mu} S_j^{\nu}.
\nonumber\\
\label{eq:extended-model}
\end{eqnarray}
The second-nearest-neighbor DM vector ${\bf D}'_{ij}$ is constrained by $C_2$ rotation symmetry as ${\bf D}'_{ij}=D'_1\hat{p}_{ij}+D'_2\hat{q}_{ij}$, where $\hat{p}_{ij}$ and $\hat{q}_{ij}$ are two orthonormal vectors perpendicular to the $C_2$ axis (see Ref. \cite{Lee2013} for the explicit expressions of $\hat{p}_{ij}$ and $\hat{q}_{ij}$).
In this calculation, we constrain the nearest-neighbor $\Gamma$ term as $\Gamma_{ij}^{\mu\nu}=\frac{D_{ij}^{\mu}D_{ij}^{\nu}}{2J_{ij}}-\frac{\delta^{\mu\nu}{\bf D}_{ij}^2}{4J_{ij}}$, and similarly for the second-nearest-neighbor $\Gamma'$ term 
(this constraint is obtained from the strong coupling expansion of a single-band Hubbard model with spin-dependent hopping channels).
Figure \ref{fig:magnon-bands} shows that the TDC and also band inversion of the original model remain robust under the additional interactions.
Furthermore, we find that this extended model shows the same sign-change behavior in $\kappa_{xy}$ as the original model does (see Fig. \ref{fig:kappa-xy-NNNmodel}).

\section{Momentum-resolved thermal Hall conductivity}

\begin{figure*}
\centering
\includegraphics[width=\linewidth]{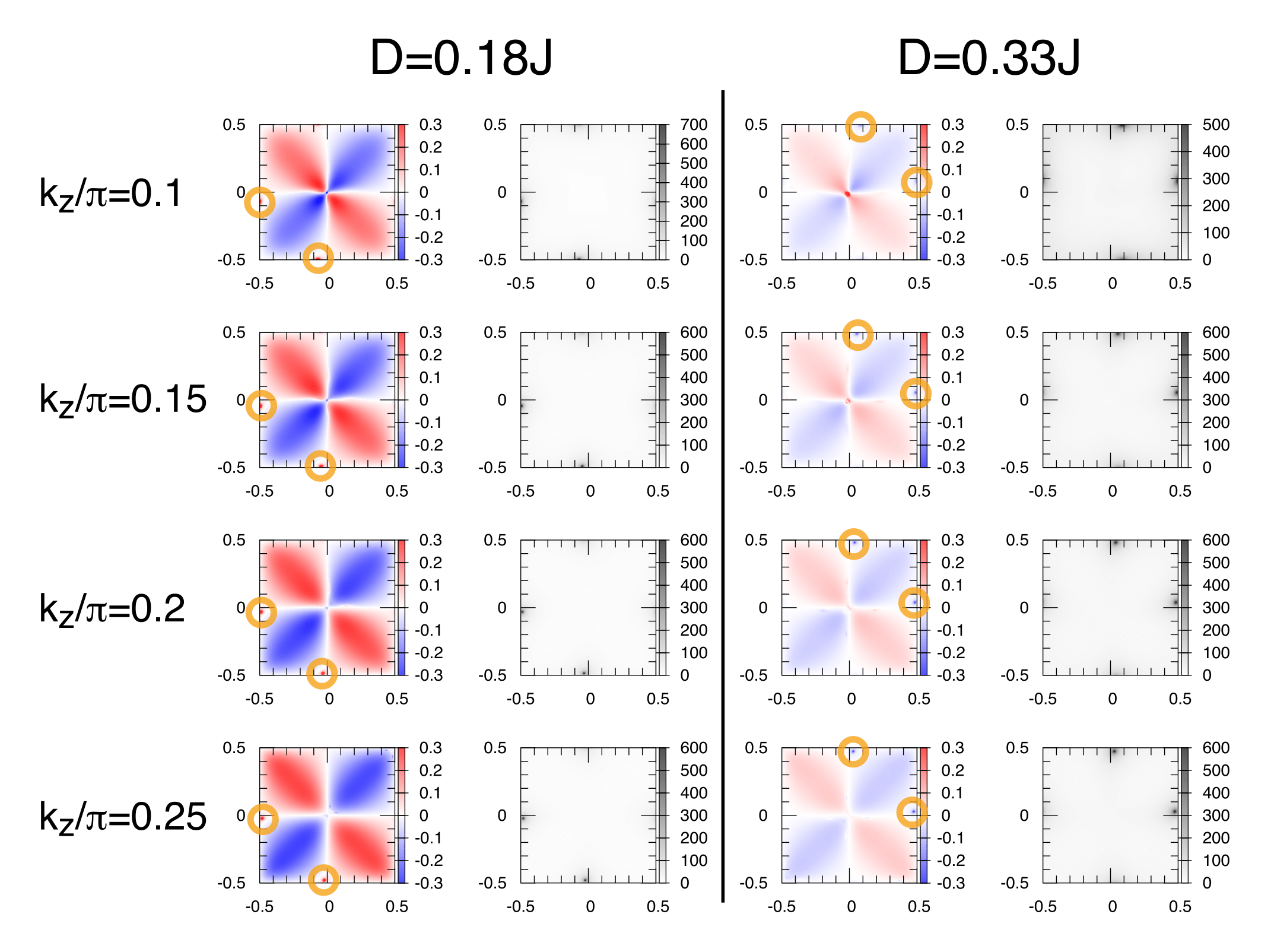}
\caption{Momentum-resolved thermal Hall conductivity $K_{xy}({\bf k})$ (color) and the inverse of the energy difference $(E_{2{\bf k}}-E_{1{\bf k}})^{-1}$ (black and white) on several $k_z$ slices of the Brillouin zone. In each plot, the horizontal and vertical axes represent $k_x/\pi$ and $k_y/\pi$, respectively. Left: $D=0.18J$. Right: $D=0.33J$.}
\label{fig:Enk-Kxy}
\end{figure*}

In the main text (Fig. 4), we have shown that the sign of $\kappa_{xy}$ is mainly determined by the (zero-field) nodal-line band crossings along XW. To convince that the high intensity peaks shown in Figs. 4 (d-e,i-j) originate from the nodal-line band crossings, we compare $K_{xy}({\bf k})$ with the inverse of the energy difference between the two lowest bands $(E_{2{\bf k}}-E_{1{\bf k}})^{-1}$. As shown in Fig. \ref{fig:Enk-Kxy}, high intensity peaks of $K_{xy}({\bf k})$ (marked by circles in the color maps) occur at the locations where $(E_{2{\bf k}}-E_{1{\bf k}})^{-1}$ is very large corresponding to the nodal-line band crossings which under an external field become nondegenerate and shifted from XW lines.

\begin{figure}
\centering
\includegraphics[width=\linewidth]{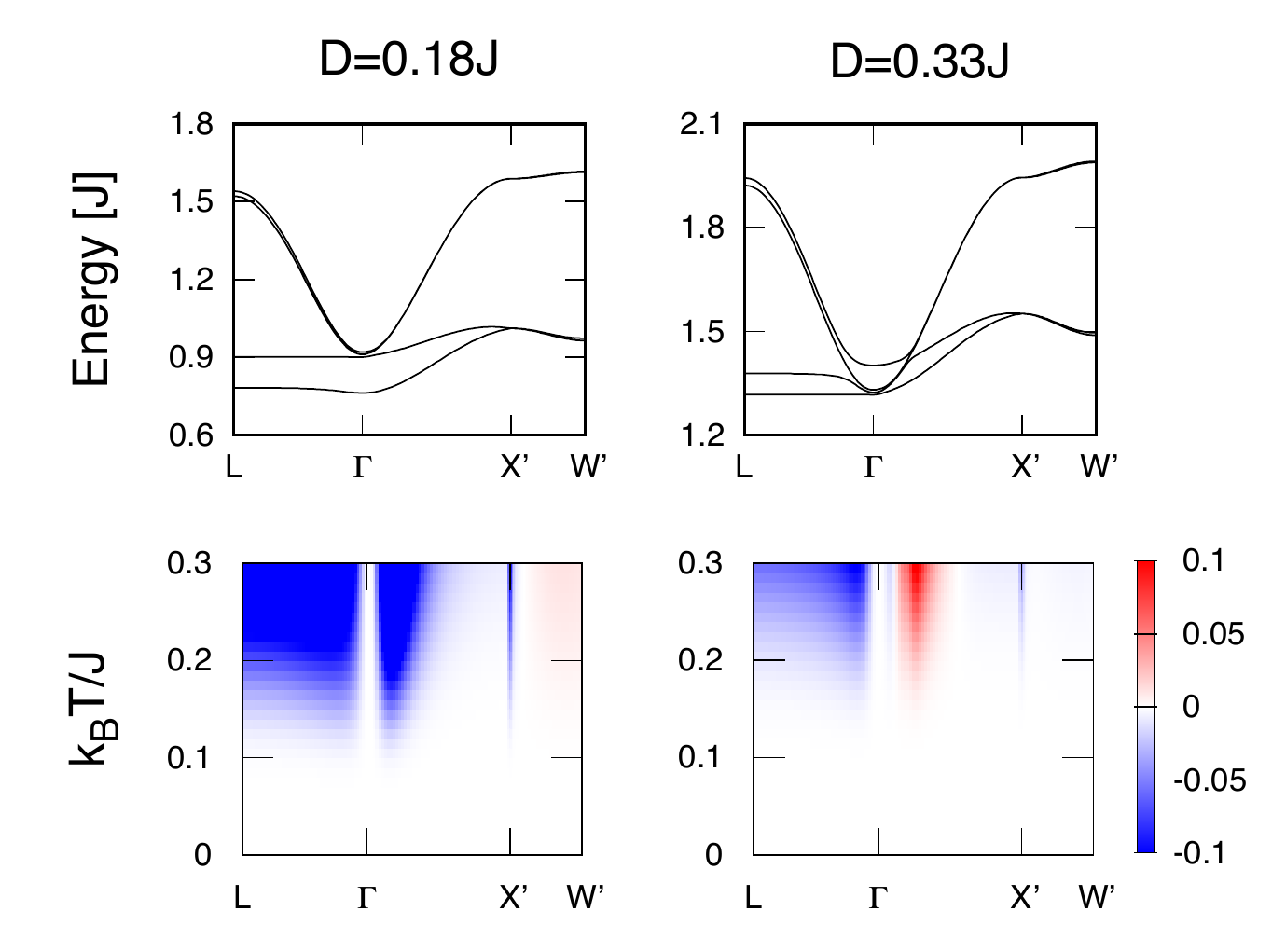}
\caption{Magnon band structure $E_{n{\bf k}}$ and momentum-resolved thermal Hall conductivity $K_{xy}({\bf k},T)$ under the external field $h=0.02J$//[110]. Left: $D=0.18J$. Right: $D=0.33J$. In each case, the color map plots $K_{xy}({\bf k},T)$ on the plane of ${\bf k}$ and $T$. In these calculations, we choose high symmetry lines connecting X$'=(0,0,\pi/2)$ and W$'=(0,\pi/4,\pi/2)$.} 
\label{fig:momres-kappa}
\end{figure}

Nodal-line band crossings arise along $\Gamma$X lines as well as XW lines [Figs. 2 (e,f)]. A natural question to ask is how large is the contribution of the former to $\kappa_{xy}$ compared to the latter. As displayed in Fig. \ref{fig:momres-kappa}, the nodal-lines along $\Gamma$X lines exhibit an opposite sign of $K_{xy}$ to that of $\kappa_{xy}$. But we find that their contribution is not large enough to cancel that of the other nodal-lines along XW lines which dominate $\kappa_{xy}$.

\begin{thebibliography}{99}

\bibitem{Vafek2014}
O. Vafek and A. Vishwanath, Annu. Rev. Condens. Matter Phys. {\bf 5}, 83 (2014).

\bibitem{Armitage2018}
N.P. Armitage, E. J. Mele, A. Vishwanath, Rev. Mod. Phys. {\bf 90}, 15001 (2018).

\bibitem{Wan2011}
X. Wan, A. M. Turner, A. Vishwanath, and S. Y. Savrasov, Phys. Rev. B {\bf 83}, 205101 (2011).

\bibitem{Burkov2011}
A. A. Burkov and Leon Balents, Phys. Rev. Lett. {\bf 107}, 127205 (2011).

\bibitem{Yang2011}
K.-Y. Yang, Y.-M. Lu, and Y. Ran, Phys. Rev. B {\bf 84}, 075129 (2011).

\bibitem{Xu2015Weyl}
S.-Y. Xu, I. Belopolski, N. Alidoust, M. Neupane, G. Bian, C. Zhang, R. Sankar, G. Chang, Z. Yuan, C.-C. Lee, S.-M. Huang, H. Zheng, J. Ma, D. S. Sanchez, B.K. Wang, A. Bansil, F. Chou, P. P. Shibayev, H. Lin, S. Jia, M. Z. Hasan, Science {\bf 349}, 613 (2015).

\bibitem{Huang2015}
S.-M. Huang, S.-Y. Xu, I. Belopolski, C.-C. Lee, G. Chang, B. Wang, N. Alidoust, G. Bian, M. Neupane, C. Zhang, S. Jia, A. Bansil, H. Lin, and M. Z. Hasan, Nat. Commun. {\bf 6}, 7373 (2015).

\bibitem{Weng2015}
H. Weng, C. Fang, Z. Fang, B. A. Bernevig, and X. Dai, Phys. Rev. X {\bf 5}, 011029 (2015).

\bibitem{Lv2015}
B. Q. Lv, H.M. Weng, B.B. Fu, X.P. Wang, H. Miao, J. Ma, P. Richard, X. C. Huang, L. X. Zhao, G. F. Chen, Z. Fang, X. Dai, T. Qian, and H. Ding, Phys. Rev. X {\bf 5}, 031013 (2015).

\bibitem{Wang2012}
 Z. Wang, Y. Sun, X.-Q. Chen, C. Franchini, G. Xu, H. Weng, X. Dai, and Z. Fang, Phys. Rev. B {\bf 85}, 195320 (2012).
 
 \bibitem{Wang2013}
 Z. Wang, H. Weng, Q. Wu, X. Dai, and Z. Fang, Phys. Rev. B {\bf 88}, 125427 (2013).

\bibitem{Yang2014_DSM}
B.-J. Yang, and N. Nagaosa, Nature Comm. {\bf 5}, 4898 (2014).
 
\bibitem{Kargarian2016}
M. Kargarian, M. Randeria, and Y.-M. Lu, 2016, PNAS {\bf 113}, 8648 (2016). 

\bibitem{Liu2014}
Z. K. Liu, J. Jiang, B. Zhou, Z. J. Wang, Y. Zhang, H. M. Weng, D. Prabhakaran, S-K. Mo, H. Peng, P. Dudin, T. Kim, M. Hoesch, Z. Fang, X. Dai, Z. X. Shen, D. L. Feng, Z. Hussain, and Y. L. Chen, Nat. Mater. {\bf 13}, 677 (2014).

\bibitem{Xu2015Dirac}
S.-Y. Xu, C. Liu, S. K. Kushwaha, R. Sankar, J. W. Krizan, I. Belopolski, M. Neupane, G. Bian, N. Alidoust,
T.-R. Chang, H.-T. Jeng, C. -Y. Huang, W. -F. Tsai, H. Lin, P. P. Shibayev, F. -C. Chou, R. J. Cava, and M. Z. Hasan, Science {\bf 347}, 294 (2015).

\bibitem{Liu2016}
Z. K. Liu, B. Zhou, Z. J. Wang, H. M. Weng, D. Prabhakaran, S. -K. Mo, Y. Zhang, Z. X. Shen, Z. Fang, X. Dai, Z. Hussain, Y. L. Chen, Science {\bf 343}, 864 (2016).

\bibitem{Zhu2016}
Z. Zhu, G. W. Winkler, Q. S. Wu, J. Li, and A. A. Soluyanov, Phys. Rev. X {\bf 6}, 031003 (2016).

\bibitem{Weng2016}
H. Weng, C. Fang, Z. Fang, and X. Dai, Phys. Rev. B {\bf 93}, 241202(R) (2016).

\bibitem{Weng2016_2}
H. Weng, C. Fang, Z. Fang, X. Dai, Phys. Rev. B {\bf 94}, 165201 (2016).

\bibitem{Bradlyn2016}
B. Bradlyn, J. Cano, Z. Wang, M. G. Vergniory, C. Felser, R. J. Cava, B. A. Bernevig, Science {\bf 353}, aaf5037 (2016).

\bibitem{Chang2017}
G. Chang, S.-Y. Xu, S.-M. Huang, D. S. Sanchez, C.-H. Hsu, G. Bian, Z.-M. Yu, I. Belopolski, N. Alidoust, H. Zheng, T.-R. Chang, H.-T. Jeng, S. A. Yang, T. Neupert, H. Lin, and M. Z. Hasan, Sci. Rep. {\bf 7}, 1688 (2017).

\bibitem{Lv2017}
B. Q. Lv, Z.-L. Feng, Q.-N. Xu, X. Gao, J.-Z. Ma, L.-Y. Kong, P. Richard, Y.-B. Huang, V. N. Strocov, C. Fang, H.-M. Weng,	 Y.-G. Shi, T. Qian, and H. Ding, Nature {\bf 546}, 627 (2017).

\bibitem{Burkov2011NLSM}
A. A. Burkov, M. D. Hook, and Leon Balents, Phys. Rev. B {\bf 84}, 235126 (2011).

\bibitem{Fang2015NLSM}
C. Fang, Y. Chen, H.-Y. Kee, and L. Fu, Phys. Rev. B {\bf 92}, 081201(R) (2015).

\bibitem{Chen2015NLSM}
Y. Chen, Y.-M. Lu, and H.-Y. Kee, Nat. Comms. {\bf 6}, 6593 (2015).

\bibitem{Chen2016NLSM}
Y. Chen, H.-S. Kim, and H.-Y. Kee, Phys. Rev. B {\bf 93}, 155140 (2016).

\bibitem{Bian2016NLSM}
G. Bian, T.-R. Chang, R. Sankar, S.-Y. Xu, H. Zheng, T. Neupert, C.-K. Chiu, S.-M. Huang, G. Chang, I. Belopolski, D. S. Sanchez, M. Neupane, N. Alidoust, C. Liu, B. Wang, C.-C. Lee, H.-T. Jeng, C. Zhang, Z. Yuan, S. Jia, A. Bansil, F. Chou, H. Lin, and M. Z. Hasan,
Nat. Comms. {\bf 7}, 10556 (2016).

\bibitem{Li2016}
F.-Y. Li, Y.-D. Li, Y. B. Kim, L. Balents, Y. Yu, and G. Chen, Nat. Comms. {\bf 7}, 12691 (2016).

\bibitem{Mook2016}
A. Mook, J. Henk, and I. Mertig, Phys. Rev. Lett. {\bf 117}, 157204 (2016).

\bibitem{Fransson2016}
J. Fransson, A. M. Black-Schaffer, A. V. Balatsky, Phys. Rev. B {\bf 94}, 075401 (2016).

\bibitem{SeKwon2016}
Se Kwon Kim, H\'ector Ochoa, Ricardo Zarzuela, and Yaroslav Tserkovnyak, Phys. Rev. Lett. {\bf 117}, 227201 (2016).

\bibitem{Owerre2016_2017_honeycomb}
S. A. Owerre, J. Phys.: Condens. Matter {\bf 28}, 386001 (2016); 
J. Phys. Commun. {\bf 1}, 025007 (2017); 
Sci. Rep. {\bf 7}, 6931 (2017).

\bibitem{Okuma2017}
N. Okuma, Phys. Rev. Lett. {\bf 119}, 107205 (2017).

\bibitem{Su2017pyrochlore}
Y. Su, X. S. Wang, and X. R. Wang, Phys. Rev. B {\bf 95}, 224403 (2017).

\bibitem{Su2017honeycomb}
Y. Su and X. R. Wang, Phys. Rev. B {\bf 96}, 104437 (2017).

\bibitem{Li2017_3Dhoneycomb}
K.-K. Li and J.-P. Hu, Chin. Phys. Lett. {\bf 34}, 077501 (2017).

\bibitem{Li2017_Cu3TeO6}
K. Li, C. Li, J. Hu, Y. Li, C. Fang, Phys. Rev. Lett. {\bf 119}, 247202 (2017).

\bibitem{Owerre2018_3Dkagome}
S. A. Owerre, Phys. Rev. B {\bf 97}, 094412 (2018).

\bibitem{Zyuzin2018}
V. A. Zyuzin and A. A. Kovalev, Phys. Rev. B {\bf 97}, 174407 (2018).

\bibitem{Jian2018}
S.-K. Jian and W. Nie, Phys. Rev. B {\bf 97}, 115162 (2018).

\bibitem{Owerre2018}
S. A. Owerre, EPL {\bf 120}, 57002 (2018).

\bibitem{Witczak-Krempa2014}
W. Witczak-Krempa, G. Chen, Y. B. Kim, and L. Balents, Annu. Rev. Condens. Matter Phys. {\bf 5}, 57 (2014).

\bibitem{Schaffer2016}
R. Schaffer, E. K.-H. Lee, B.-J. Yang, and Y. B. Kim, Rep. Prog. Phys. {\bf 79}, 094504 (2016).

\bibitem{Yanagishima2001}
D. Yanagishima and Y. Maeno, J. Phys. Soc. Jpn. {\bf 70}, 2880 (2001).

\bibitem{Taira2001}
N. Taira, M. Wakeshima, and Y. Hinatsu, J. Phys.: Condens. Matter {\bf 13}, 5527 (2001).

\bibitem{Fukazawa2002}
H. Fukazawa and Y. Maeno, J. Phys. Soc. Jpn. {\bf 71}, 2578 (2002).

\bibitem{Matsuhira2007}
K. Matsuhira, M. Wakeshima, R. Nakanishi, T. Yamada, A. Nakamura, W. Kawano, S. Takagi, and Y. Hinatsu, J. Phys. Soc. Jpn. {\bf 76}, 043706 (2007).

\bibitem{Hasegawa2010}
T. Hasegawa, N. Ogita, K. Matsuhira, S. Takagi, M. Wakeshima, Y. Hinatsu, and M. Udagawa, J. Phys.: Conf. Ser. {\bf 200}, 012054 (2010).

\bibitem{Machida2010}
Y. Machida, S. Nakatsuji, S. Onoda, T. Tayama, and T. Sakakibara, Nature {\bf 463}, 210 (2010).

\bibitem{Sakata2011}
M. Sakata, {\it et al.}, Phys. Rev. B {\bf 83}, 041102 (2011).

\bibitem{Zhao2011}
S. Zhao, J. M. Mackie, D. E. MacLaughlin, O. O. Bernal, J. J. Ishikawa, Y. Ohta, and S. Nakatsuji, Phys. Rev. B {\bf 83}, 180402(R) (2011).

\bibitem{Matsuhira2011}
K. Matsuhira, M. Wakeshima, Y. Hinatsu, and S. Takagi, J. Phys. Soc. Jpn. {\bf 80}, 094701 (2011).

\bibitem{Tomiyasu2012}
K. Tomiyasu, K. Matsuhira, K. Iwasa, M. Watahiki, S. Takagi, M. Wakeshima, Y. Hinatsu, M. Yokoyama, K. Ohoyama, and K. Yamada, J. Phys. Soc. Jpn. {\bf 81}, 034709 (2012).

\bibitem{Disseler2012}
S. M. Disseler, C. Dhital, T. C. Hogan, A. Amato, S. R. Giblin, C. de la Cruz, A. Daoud-Aladine, S. D. Wilson, and M. J. Graf, Phys. Rev. B {\bf 85}, 174441 (2012).

\bibitem{Disseler2012_II}
S. M. Disseler, C. Dhital, A. Amato, S. R. Giblin, C. de la Cruz, S. D. Wilson, and M. J. Graf, Phys. Rev. B {\bf 86}, 014428 (2012).

\bibitem{Tafti2012}
F. F. Tafti, J. J. Ishikawa, A. McCollam, S. Nakatsuji, and S. R. Julian, Phys. Rev. B {\bf 85}, 205104 (2012).

\bibitem{Shapiro2012}
M. C. Shapiro, S. C. Riggs, M. B. Stone, C. R. de la Cruz, S. Chi, A. A. Podlesnyak, and I. R. Fisher, Phys. Rev. B {\bf 85}, 214434 (2012).

\bibitem{Ishikawa2012}
J. J. Ishikawa, E. C. T. O'Farrell, and S. Nakatsuji, Phys. Rev. B {\bf 85}, 245109 (2012).

\bibitem{Sagayama2013}
H. Sagayama, D. Uematsu, T. Arima, K. Sugimoto, J. J. Ishikawa, E. O'Farrell, and S. Nakatsuji, Phys. Rev. B {\bf 87}, 100403 (2013).

\bibitem{Guo2013}
H. Guo, K. Matsuhira, I. Kawasaki, M. Wakeshima, Y. Hinatsu, I. Watanabe, and Z.-a. Xu, Phys. Rev. B {\bf 88}, 060411 (2013).

\bibitem{Kondo2015}
T. Kondo, M. Nakayama, R. Chen, J. J. Ishikawa, E.-G. Moon, T. Yamamoto, Y. Ota, W. Malaeb, H. Kanai, Y. Nakashima, Y. Ishida, R. Yoshida, H. Yamamoto, M. Matsunami, S. Kimura, N. Inami, K. Ono, H. Kumigashira, S. Nakatsuji, L. Balents, S. Shin, Nature comm. {\bf 6}, 10042 (2015).

\bibitem{Ueda2015}
K. Ueda, J. Fujioka, B.-J. Yang, J. Shiogai, A. Tsukazaki, S. Nakamura, S. Awaji, N. Nagaosa, and Y. Tokura, Phys. Rev. Lett. {\bf 115}, 056402 (2015).

\bibitem{Tian2016}
Z. Tian, Y. Kohama, T. Tomita, H. Ishizuka, T. H. Hsieh, J. J. Ishikawa, K. Kindo, L. Balents, and S. Nakatsuji, Nature Phys. {\bf 12}, 134 (2016).

\bibitem{Clancy2016}
J. P. Clancy, H. Gretarsson, E. K. H. Lee, D. Tian, J. Kim, M. H. Upton, D. Casa, T. Gog, Z. Islam, B.-G. Jeon, K. H. Kim, S. Desgreniers, Y. B. Kim, S. J. Julian, Y.-J. Kim, Phys. Rev. B {\bf 94}, 024408 (2016).

\bibitem{Donnerer2016}
C. Donnerer, M. C. Rahn, M. M. Sala, J. G. Vale, D. Pincini, J. Strempfer, M. Krisch, D. Prabhakaran, A. T. Boothroyd, and D. F. McMorrow, Phys. Rev. Lett. {\bf 117}, 037201 (2016).


\bibitem{fujita2015}
T. Fujita, Y. Kozuka, M. Uchida, A. Tsukazaki, T. Arima, and M. Kawasaki, Sci. Rep. {\bf 5} (2015).

\bibitem{fujita2016}
T. Fujita, M. Uchida, Y. Kozuka, S. Ogawa, A. Tsukazaki, T. Arima, and M. Kawasaki, Appl. Phys. Lett. {\bf 108}, 022402 (2016).

\bibitem{fujita2016_2nd}
T. C. Fujita, M. Uchida, Y. Kozuka, W. Sano, A. Tsukazaki, T. Arima, and M. Kawasaki, Phys. Rev. B {\bf 93}, 064419 (2016).

\bibitem{Witczak-Krempa2012} W. Witczak-Krempa and Y. B. Kim, Phys. Rev. B {\bf 85}, 045124 (2012).

\bibitem{Go2012}
A. Go, W. Witczak-Krempa, G. S. Jeon, K. Park, and Y. B. Kim, Phys. Rev. Lett. {\bf 109}, 066401 (2012).

\bibitem{Moon2013}
E.-G. Moon, C. Xu, Y. B. Kim, and L. Balents, Phys. Rev. Lett. {\bf 111}, 206401 (2013).

\bibitem{Lee2013}
E. K.-H. Lee, S. Bhattacharjee, and Y. B. Kim, Phys. Rev. B {\bf 87}, 214416 (2013).

\bibitem{Chen2012}
G. Chen and M. Hermele, Phys. Rev. B {\bf 86}, 235129 (2012).

\bibitem{Chen2015}
Q. Chen, H.-H. Hung, X. Hu, and G. A. Fiete, Phys. Rev. B {\bf 92}, 085145 (2015).

\bibitem{Zhang2017}
H. Zhang, K. Haule, and D. Vanderbilt, Phys. Rev. Lett. {\bf 118}, 026404 (2017).

\bibitem{Wang2017}
R. Wang, A. Go, and A. J. Millis, Phys. Rev. B {\bf 95}, 045133 (2017).

\bibitem{hu2012}
X. Hu, A. R\"uegg, and G. A. Fiete, Phys. Rev. B {\bf 86}, 235141 (2012).

\bibitem{Yang2014_WSM} 
B.-J. Yang and N. Nagaosa, 
Phys. Rev. Lett. {\bf 112}, 246402 (2014).

\bibitem{yamaji2014}
Y. Yamaji and M. Imada, Phys. Rev. X {\bf 4}, 021035 (2014).

\bibitem{hu2015}
X. Hu, Z. Zhong, and G. A. Fiete, Sci. Rep. {\bf 5}, 11072 (2015).

\bibitem{Hwang2016}
K. Hwang and Y. B. Kim, Sci. Rep. {\bf 6}, 30017 (2016).

\bibitem{Laurell2017}
P. Laurell and G. A. Fiete, Phys. Rev. Lett. {\bf 118}, 177201 (2017). 


\bibitem{Katsura2010}
H. Katsura, N. Nagaosa, and P. A. Lee, Phys. Rev. Lett. {\bf 104}, 066403 (2010).

\bibitem{Onose2010}
Y. Onose, T. Ideue, H. Katsura, Y. Shiomi, N. Nagaosa, and Y. Tokura, Science {\bf 329}, 297 (2010).

\bibitem{Ideue2012}
T. Ideue, Y. Onose, H. Katsura, Y. Shiomi, S. Ishiwata, N. Nagaosa, and Y. Tokura
Phys. Rev. B {\bf 85}, 134411 (2012).

\bibitem{Matsumoto2011}
R. Matsumoto and S. Murakami, Phys. Rev. Lett. {\bf 106}, 197202 (2011);
R. Matsumoto and S. Murakami, Phys. Rev. B {\bf 84}, 184406 (2011);

\bibitem{Matsumoto2014}
R. Matsumoto, R. Shindou, and S. Murakami, Phys. Rev. B {\bf 89}, 054420 (2014).

\bibitem{Hirschberger2015}
M. Hirschberger, R. Chisnell, Y. S. Lee, and N. P. Ong, Phys. Rev. Lett. {\bf 115}, 106603 (2015).

\bibitem{Lee2015}
H. Lee, J. H. Han, and P. A. Lee, Phys. Rev. B {\bf 91}, 125413 (2015).

\bibitem{Elhajal2005}
M. Elhajal, B. Canals, R. Sunyer, and C. Lacroix, Phys. Rev. B {\bf 71}, 094420 (2005).

\bibitem{suppl}
See Supplemental Material for details of spin wave theory, symmetry analysis, magnon band structure, thermal Hall conductivity calculations, and influence of further-neighbor interactions. 

\bibitem{Rau2016}
J. G. Rau, E. K.-H. Lee, and H.-Y. Kee, Annu. Rev. Condens. Matter Phys. {\bf 7}, 195 (2016).

\end{thebibliography}

\begin{thebibliography}{99}

\bibitem{Holstein1940}
T. Holstein and H. Primakoff, Phys. Rev. {\bf 58}, 1098 (1940).

\bibitem{Onose2010}
Y. Onose, T. Ideue, H. Katsura, Y. Shiomi, N. Nagaosa, and Y. Tokura, Science {\bf 329}, 297 (2010).

\bibitem{Ideue2012}
T. Ideue, Y. Onose, H. Katsura, Y. Shiomi, S. Ishiwata, N. Nagaosa, and Y. Tokura
Phys. Rev. B {\bf 85}, 134411 (2012).

\bibitem{Matsumoto2014}
R. Matsumoto, R. Shindou, and S. Murakami, Phys. Rev. B {\bf 89}, 054420 (2014).

\bibitem{Soluyanov2015}
A. A. Soluyanov, D. Gresch, Z. Wang, Q. Wu, M. Troyer, X. Dai, and B. A. Bernevig, Nature {\bf 527}, 495 (2015).

\bibitem{Li2016}
F.-Y. Li, Y.-D. Li, Y. B. Kim, L. Balents, Y. Yu, and G. Chen, Nat. Comms. {\bf 7}, 12691 (2016).

\bibitem{Mook2016}
A. Mook, J. Henk, and I. Mertig, Phys. Rev. Lett. {\bf 117}, 157204 (2016).

\bibitem{Lee2013}
E. K.-H. Lee, S. Bhattacharjee, and Y. B. Kim, Phys. Rev. B {\bf 87}, 214416 (2013).

\end{thebibliography}
\end{document}